\documentclass[useAMS,usenatbib]{mnras}

\usepackage{graphicx}\usepackage{epsfig}\usepackage{epsf}
\usepackage{amsmath}\usepackage{amssymb}\usepackage{stfloats}
\newcommand{\Msun}{\,$M_{\odot}$}
\newcommand{\Rsun}{\,$R_{\odot}$}
\newcommand{\Lsun}{\,$L_{\odot}$}
\newcommand{\kms}{\,km\,s$^{-1}$}
\newcommand{\Msunyr}{\,$M_{\odot}$\,yr$^{-1}$}
\newcommand{\rev}[1]{{{\bf #1}}}  
\usepackage{booktabs}    

\def\bain{\ {Bul. Astron. Inst. Neth.}\ }
\usepackage[dvipsnames]{xcolor}
          
\title[Isolated WN3/O3 stars]{Extreme isolation of WN3/O3 stars and
  implications for their evolutionary origin as the elusive stripped binaries}

\author[Smith, G\"{o}tberg, \& de Mink]{Nathan
  Smith$^{1,3}$\thanks{E-mail: nathans@as.arizona.edu}, Ylva
  G\"otberg$^{2,3}$, and Selma E.\ de Mink$^{2,3}$ \\ $^{1}$Steward
  Observatory, University of Arizona, 933 N. Cherry Ave., Tucson, AZ
  85721, USA \\ $^2$Anton Pannekoek Institute for Astronomy,
  University of Amsterdam, Science Park 904, 1098XH Amsterdam, The Netherlands\\
  $^{3}$Kavli Institute for Theoretical Physics, University of California, Santa
Barbara, CA 93106, USA 
  }

\begin{document}

\pagerange{\pageref{firstpage}--\pageref{lastpage}} \pubyear{2017}
\maketitle
\label{firstpage}

\begin{abstract}

  Recent surveys of the Magellanic Clouds have revealed a subtype
  of Wolf-Rayet (WR) star with peculiar properties. WN3/O3 spectra
  exhibit both WR-like emission and O3 V-like absorption --- but at
  lower luminosity than O3 V or WN stars.  We examine the
  projected spatial distribution of WN3/O3 stars in the LMC as
  compared to O-type stars.  Surprisingly, WN3/O3 stars are among the most
  isolated of all classes of massive stars; they have a distribution
  similar to red supergiants dominated by initial masses
  of 10-15\Msun, and are far more dispersed than classical WR stars
  or luminous blue variables (LBVs).  Their lack of association with clusters 
  of O-type stars suggests strongly that WN3/O3 stars are not the
  descendants of single massive stars (30\Msun \, or above).
  Instead, they are likely products of interacting binaries at
  lower initial mass (10-18\Msun). Comparison with binary models suggests a
  probable origin with primaries in this mass range
  that were stripped of their H envelopes through non-conservative mass
  transfer by a low-mass secondary.  
  We show that model spectra and positions on the Hertzsprung-Russell diagram 
  for binary stripped stars are consistent with WN3/O3 stars.  Monitoring 
  radial velocities with high-resolution spectra can test for 
  low-mass companions or runaway velocities.  With lower initial 
  mass and environments that avoid very massive stars,
  the WN3/O3 stars fit expectations for progenitors of Type~Ib and
  possibly Type~Ibn supernovae.

\end{abstract}

\begin{keywords}
  stars: evolution --- stars: massive --- stars: winds, outflows ---
  stars: Wolf-Rayet
\end{keywords}

\section{INTRODUCTION \label{intro}}

Wolf-Rayet (WR) stars represent a late evolutionary phase when massive
stars have been stripped of their H envelopes, exposing the He
core (see \citealt{crowther07} for a review). Most WR stars come in
two main varieties, with spectra dominated by either nitrogen (WN
stars) or carbon (WC stars). A third type, WO stars, are the most
stripped but are extremely rare. (We exclude the WNH
stars, which are H-rich and likely near the end of the 
main-sequence; see \citealt{sc08}.)  The two dominant mechanisms proposed for 
removing the H envelopes to make H-poor WR stars are stellar winds
or binary mass transfer.  Both may be at work, and their contributions
to WR populations may differ with initial mass and metallicity.

The standard single-star view, often referred to as the ``Conti
scenario'' \citep{conti76}, holds that WR stars are the descendants of
the most massive O-type stars by virtue of their own mass loss.
Powerful radiation-driven winds act over the lifetime of the star to
drive off its outer layers, so that all stars above some initial mass
threshold (35\,\Msun, for example, but this depends on
adopted mass-loss prescriptions; \citealt{renzo17}) will 
end their lives as WR stars.  In this view, all stars above the threshold 
will yield WN stars first, but only stars with the highest luminosities
produce WC or WO stars \citep{groh14}.  For lower initial mass where the 
winds are weak, stars die as H-rich red supergiants (RSGs) and
Type II-P supernovae (SNe), with a narrow range in between
exploding in a transitional luminous blue variable (LBV) phase
\citep{groh14}.

Over the years, the Conti scenario has found support in comparisons
between model predictions and the observed number ratios of different
types of stars in nearby galaxies \citep{meynet11,mm03,massey03}.
However, some considerations have begun to unravel this paradigm as
the dominant mechanism controlling the evolution of massive stars.
Recent estimates of mass-loss rates have been revised downward,
lower than mass-loss rates adopted in most models (see
\citealt{smith14} for a recent review). These weaker winds would then
require the assistance of eruptive mass loss from LBVs to
shed their H envelopes and make WR stars
\citep{so06}. However, LBVs cannot participate in this evolutionary
sequence of single massive stars as envisioned in models, since
their isolated environments require that they are primarily
products of binary interaction \citep{st15}.

A different explanation for the origin of WR stars acknowledges that
interacting binary systems exist in large numbers and profoundly
influence evolution \citep{paczynski71}. Mass transfer in
interacting binaries can operate over a wider initial mass range than strong 
winds, stripping the H envelope in stars that are not luminous enough to do so 
with their own mass loss or at low metallicity \citep{2017A&A...608A..11G}.  The large 
binary fraction dictates that mass transfer and mergers will dominate the
observed statistics of massive stars \citep{sana12,Kobulnicky+2014,2016arXiv160605347M}.  
Statistics of the observed fractions of SN subtypes \citep{smith11} as 
well as the properties derived for individual stripped-envelope SNe
\citep{drout11,dessart11,haschinger12,2015ApJ...809..131K,lyman16,yoon17}
strongly favor the interpretation that most stripped envelope SNe (Types 
Ic, Ib, IIb) arise from relatively low initial mass progenitors stripped 
in binaries.  Since very massive stars might in
some cases collapse to a black hole without a bright SN display, it
remains unclear what fraction of stripped-envelope SNe arise
from classical single WR star progenitors.

In an effort to have a more complete census of WR stars in nearby 
environments, Massey and collaborators have been conducting a survey 
of the Magellanic Clouds (LMC/SMC) with modern interference-filter 
imaging \citep{massey14,massey15,massey17}.  Among the results is the 
reported discovery of a peculiar category of WR stars called WN3/O3 stars 
in the LMC. Similar to most of the WR stars in the SMC, these LMC stars 
have spectra that show composite properties with emission lines of early 
WN stars, but also absorption features of O3~V stars.  However, Massey 
et al.\ point out that these are far too faint to be simple composite 
spectra of binary WN + O3~V systems. Such a pairing also seems unlikely, 
since the lifetimes of O3~V stars are too short to allow a companion to evolve
to the WN phase \citep{massey14}. Massey et al.\ therefore conclude
that this is a new type of WR star, not simply two stars contributing
to the spectrum.  Comparison of WN3/O3 spectra with model atmosphere
codes points to high photospheric temperatures (80,000-100,000~K), a
factor of 10 enhancement in N abundance, H depletion, luminosities
around 10$^5$\Lsun, and relatively low mass-loss rates of
$\sim$10$^{-6}$\Msunyr \citep{massey14,neugent16,neugent17}. In the $V$ 
band, WN3/O3 stars are a few magnitudes fainter than classical WN stars 
\citep{crowther07}. Their evolutionary origin is uncertain, as such stars 
are not expected at these low luminosities from single-star evolution 
at LMC metallicity.

The local environments of evolved massive stars and SNe can provide
important clues about their lifetimes, and may provide insight to
their previous evolutionary path. Typically, the WN3/O3 stars
discovered so far do not appear in dense clustered environments where
the most massive stars reside \citep{massey14,massey15,massey17}.
\citet{massey15} suggest that at first glance, however, there is
nothing special or unusual about the spatial distributions of WN3/O3
stars as compared to classical WR stars, since they are scattered across
similar regions of the LMC (see their Figure 8). \citet{massey15}
therefore conclude that their environments do not point to a different
progenitor population as compared to normal WN stars.

In this paper we first take a closer look at the environments of WN3/O3
stars. In particular, we use the same metric used previously by \citet{st15} 
to investigate the environments of LBVs, by measuring the projected separation on 
the sky to the nearest known O-type stars. This provides a diagnostic of the 
relative age in the near vicinity.  The main result of this analysis
is that the WN3/O3 stars are surprisingly dispersed on the sky,
showing no correlation with O-type stars, and with a spatial distribution 
similar to RSGs.  This limits their initial masses to be below 18\Msun 
(that of late O-type main sequence stars), and provides strong 
implications for their previous evolution.

We therefore compare the WN3/O3 stars with model predictions including single and binary star evolution.  In particular, we consider new atmosphere models for stars that have been stripped of their H envelope through binary evolution \citep{2017A&A...608A..11G}, showing that the characteristic mix of absorption and emission features in WN3/O3 stars is naturally reproduced by these models (Gotberg et al. submitted). Altogether, we find that the spectral features, the remote location, their low luminosities with respect to other WR stars, and their rate can all be readily explained if WN3/O3 stars belong to the theoretically predicted class of elusive He stars that formed through Roche-lobe overflow in a binary system.

\begin{center}
\begin{table}
\label{tab:list}
\begin{minipage}{3.3in}%
    \caption{LMC WN3/O3 stars and the projected separation (S) to the
      nearest known O-type star}\scriptsize
\begin{tabular}{@{}lccccc}\hline\hline
Name      &Ref.  &$\alpha_{2000}$  &$\delta_{2000}$  &M$_V$ (mag) &$S$ (deg)  \\ \hline
LMC079-1  &(1) &05 07 13.33 &-70 33 33.9 &$-$2.8  &0.412 \\
LMC170-2  &(1) &05 29 18.18 &-69 19 43.2 &$-$2.9  &0.273 \\
LMC172-1  &(1) &05 35 00.90 &-69 21 20.2 &$-$3.0  &0.103 \\
LMC174-1  &(1) &05 40 03.57 &-69 37 53.1 &$-$3.0  &0.016 \\
LMC199-1  &(1) &05 28 27.12 &-69 06 36.2 &$-$2.3  &0.117 \\
LMC277-2  &(1) &05 04 32.65 &-68 00 59.7 &$-$3.1  &0.187 \\
LMCe159-1 &(2) &05 24 56.89 &-66 26 44.5 &$-$2.6  &0.735 \\
LMCe169-1 &(2) &05 21 22.84 &-65 52 49.0 &$-$1.8  &0.394 \\
LMCe078-3 &(3) &05 41 17.50 &-69 06 56.2 &$-$2.2  &0.151 \\
LMCe055-1 &(3) &04 56 48.72 &-69 36 40.3 &$-$2.8  &0.455 \\
\hline
\end{tabular}
\medskip
$S$ is the projected separation on the sky in degrees between an WN3/O3 star and its nearest known O-type star neighbor of any subtype or luminosity class.  LMCe055R-1 is a WN4/O4 type, but we include it here as noted in the text.  References for discovery, $M_V$, and coordinates: (1) \citet{massey14}; (2) \citet{massey15}; (3) \citet{massey17}.\\
\end{minipage}\end{table}
\end{center}

\begin{figure*}
\includegraphics[width=5.5in]{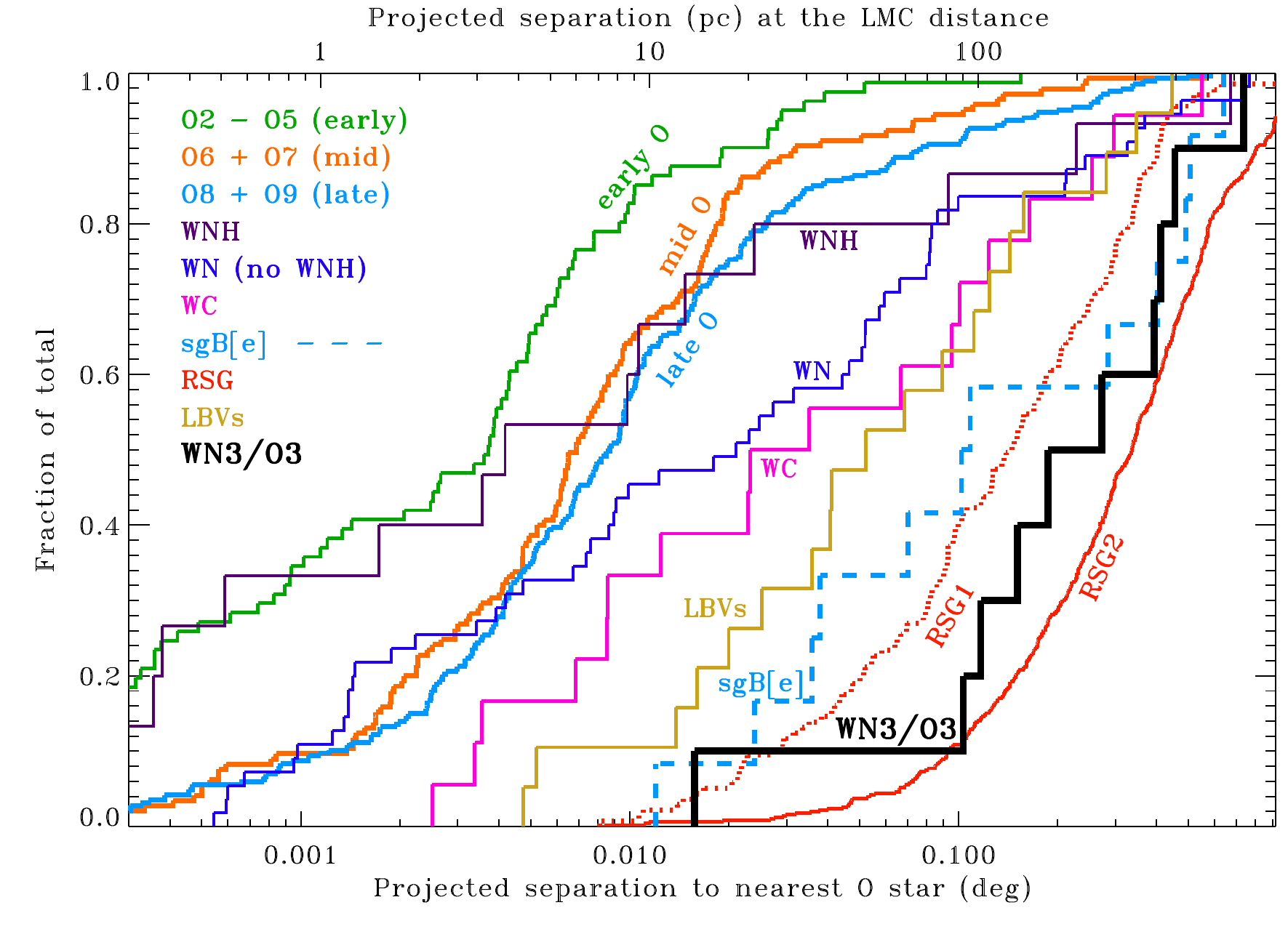}
\caption{A cumulative distribution plot of the projected separation on
  the sky between various types of massive stars and their nearest
  catalogued O-type neighbor in the LMC.  This is very similar to Figure 4 of
  \citet{st15}, except that we now include WN3/O3
  stars, and we also separate WNH and classical WN stars as noted
  previously by \citet{smith16}. This plot also uses an updated list of comparison O stars, and includes two separate samples of RSG stars (RSG1 and RSG2; see text).  See \citet{st15} for further details.}
\label{fig:cumplot}
\end{figure*}

\section{OBSERVED ENVIRONMENTS OF WN3/O3 STARS \label{isolation}}

So far, ten WN3/O3 stars have been discovered in the LMC 
\citep{massey14,massey15,massey17}.  (We include the WN4/O4 star 
LMCe055-1 in this class, even though it is slightly cooler.)  These 
ten objects with absolute magnitudes ($M_V$) and coordinates are listed 
in Table~\ref{tab:list}.

To help constrain the likely evolutionary history of WN3/O3 stars, we 
adopt the same metric used previously by \citet{st15} to quantify the 
proximity of young massive stars in the environments of LBVs.
The analysis is to measure the projected separation on the sky ($S$ in
degrees) between individual stars of a certain class and their nearest
known O-type star neighbor, and to then compare the distribution of these
separations to those of other types of massive stars.  We measure the
distance to the nearest catalogued O star of any subtype, which is dominated 
by later-type O8-O9 stars; this provides the best comparison for all but
the most massive stars, where a comparison restricted to only early
O types would be appropriate \citep{smith16}.  \citet{st15} calculated
values of $S$ for LBVs in the LMC and SMC, and compared that
distribution to the distributions of $S$ for early, mid, and late
O-type stars, WR stars and their subtypes, RSGs, and B[e] supergiants.
By comparing the separations of LBVs, O-type stars, and WR stars,
\citet{st15} demonstrated that LBVs cannot be a transitional stage
between O stars and WR stars as envisioned in the single-star paradigm, 
and concluded that they must be the product of interacting binaries,
either by rejuvenation through mass transfer or mergers, or by
receiving a kick from a companion's SN explosion.

We adopt the same technique and comparison data here, so we refer the
reader to \citet{st15} for further details.  The basic idea is that
O-type stars are mostly born in clusters and associations, so that the
average observed separation between O stars starts out small, but
slowly grows with time as unbound clusters drift apart and as the most
massive stars die off and remove themselves from the observed
population.  The most massive stars have short main-sequence lifetimes
and small average separations, so their descendants must also have
relatively small separations, only slightly larger than for early or
mid O-type stars. Evolved stars that descend from lower initial masses
(late O-type or early B-type stars, for example) are skewed to larger
separations.  Different evolutionary stages that are sequential should
move in-step from left to right on Figure~\ref{fig:cumplot}.  Adopting
a simple model for the passive dissolution of star clusters,
\citet{mojgan17} demonstrated that the observed distributions of
early, mid, and late-type O stars from \citet{st15} could be explained
quantitatively with a secular drift velocity of 10-20\kms and
the expected ages of single O-type stars.  They confirmed that the
observed separations of LBVs, however, cannot be explained by the same
model of drifting single stars, so that LBVs require longer lifetimes
than allowed for single stars of their observed luminosity, or
significant kicks.

We measured the projected separation on the sky to the nearest known
O-type star (of any subtype) for the ten known WN3/O3 stars, and these
values are listed in Table~\ref{tab:list}.
Figure~\ref{fig:cumplot} shows the cumulative distribution of these
$S$ values in black, as compared to distributions for other types of
massive stars including early, mid, and late O-type stars, WNH, WN, WC,
LBVs, B[e] supergiants, and RSGs.  Figure~\ref{fig:cumplot} is very similar 
to plots presented by \citet{st15}, except that we have added WN3/O3 stars 
of  course, plus a few other modifications.  We divided the other WR 
subtypes slightly differently, separating WNH and WN stars as explained
by \citet{smith16}, we have updated the list of O-type comparison stars 
(generated the same way as \citealt{st15}), and we now include two samples 
of RSGs, RSG1 and RSG2.

Since the distribution of WN3/O3 separations turns out to be similar to RSGs, we have considered these in more detail.  RSG1 (dashed red line) is the sample of RSGs from \citet{st15}. These were selected from known stars in SIMBAD based on their luminosity and spectral type (chosen to avoid lower-mass AGB stars), favoring fairly luminous RSGs that sample initial masses of 12-18 $M_{\odot}$ for standard non-rotating models, or more like 10-15 $M_{\odot}$ for rotating models. RSG2 (solid red) is a more complete sample of RSGs that includes lower-mass stars.  This list was provided to us by M.\ Drout (private comm.), with membership refined from the sample discussed by \citet{2012ApJ...749..177N}.  Based on Figure 6 in that paper, this RSG2 sample is dominated by lower-luminosity RSGs mostly in the initial mass range of 8--12 $M_{\odot}$.  One can therefore think of RSG1 as a sample of more massive RSGs of 12-18 $M_{\odot}$, and RSG2 as including mostly older and lower-mass RSGs of 8-12 $M_{\odot}$.  As expected, the older group RSG2 is skewed farther to the right than RSG1 in Fig~\ref{fig:cumplot}.  It is informative that the WN3/O3 stars reside in between RSG1 and RSG2 in this plot.  This is a valuable independent verification that uses bright RSGs as a reference, avoiding any potential incompleteness at the fainter end of the O star comparison sample (discussed in the next section).  This is a strong indication that the WN3/O3 stars have relatively low initial masses similar to common RSGs.

The end result in Figure~\ref{fig:cumplot} is quite surprising ---
{\it the new class of WN3/O3 stars is among the most dispersed populations
  of known types of evolved massive stars}, and it is significantly more
isolated than classical WR stars.  Whereas WNH stars have a median
separation of $\sim$4\,pc, and H-poor WN stars have a median separation
of $\sim$20\,pc, the WN3/O3 stars have a median separation from O-type
stars of $\sim$200\,pc.  Such large separations are similar to the tail
end of the maximum separations between late O-type stars, and as such,
the projected separations between WN3/O3 stars and O stars may be
consistent with a chance projection among the field O stars.  In
other words, the nearest neighboring O-type stars seen in projection
may have little to do with the actual birth population of WN3/O3
stars; O star siblings of the WN3/O3 stars may be long-since
dead.  This implies that the average initial mass of WN3/O3 stars is less 
than that of the lowest-mass O-type stars (about 18 $M_{\odot}$), although 
there may be a range.  For example, one of the WN3/O3 stars (LMC174-1) is 
indeed in a clustered region with mid- and late-O type stars, but the rest 
of the objects appear to avoid O-star clusters.  As noted above, WN3/O3 
stars are similarly isolated from O-type stars as
RSGs in the LMC.  This gives a strong indication that the WN3/O3 stars have 
long lifetimes commensurate with H-burning main sequence ages 
at relatively low initial masses of $\sim$15\Msun.

Their progenitors appear to be a different population from those of the
classical WN and WC stars, which are usually thought to arise from stars 
of $\sim$30\Msun\ or more (of course, many of the classical WR stars could arise from binary mass transfer as well).  For example, a Kolmogorov-Smirnov 
(KS) test to quantify the significance of the difference in O-star separation
distributions for normal WN stars and WN3/O3 stars yields a $p$-value
of 7.4$\times$10$^{-5}$; i.e. there is little chance that normal
WN stars and WN3/O3 stars are drawn from the same parent distribution.
In contrast, a KS test comparing the WN3/O3 distribution to that of
RSGs yields $p$-values of 0.25 and 0.6 for RSG1 and RSG2, respectively, 
consistent with these groups arising from parent distributions with similar 
ages.  In other words, if the  WN3/O3 stars had been born as single stars, 
they would most likely be in the RSG population.  

Single stars with initial masses around 15 $M_{\odot}$ do not
have strong enough mass loss to shed their H envelopes via winds.
WN3/O3 stars are therefore most likely the product of mass
stripping in binaries, and we discuss the evolutionary implications in 
Section~\ref{evolutionary_origin}.

\section{POSSIBLE SELECTION EFFECTS \label{selection}}

With any analysis of a cumulative distribution, one must consider
selection effects and how completeness of the comparison samples may
influence the results.  The survey by Massey et al.\ that revealed the
existence of the WN3/O3 stars is still underway, so additional WN3/O3
stars may yet be out there.  This is immaterial, though, since our
study investigates the local environment for each star by documenting
the nearest known O star; their overall distribution within the host galaxy
is not considered.  

Within the surveyed area, Massey et al.\ argue that their detection of
WN3/O3 stars is complete, noting that their 5$\sigma$ detection limit
was at least a magnitude fainter than the observed WN3/O3 stars.
\citet{massey15} took this as indicating that they are not
missing a fainter population of WN3/O3 stars.  If the WN3/O3 sample 
is complete in the surveyed area, what about the comparison O stars?
Indeed, it is reasonable to wonder if the census of late O8-O9 dwarfs
in the LMC, which dominate the field O stars, may be
incomplete.  As noted by \citet{st15}, though, we can ask what effect
it would have if we randomly peppered the LMC field with additional
O-type stars.  In that case, WN3/O3 stars might shift slightly to the
left in Figure~\ref{fig:cumplot}, but importantly, so would the RSGs, 
B[e] supergiants, LBVs and WR stars, because they would all be more 
likely to have a slightly nearer O-type neighbor as well. As such, WN3/O3
stars would remain relatively isolated as compared to other types of
evolved massive stars.  O stars provide the large background sample to 
which all the other types of evolved stars are compared, so incompleteness 
in that sample does not explain why WN3/O3 stars are so isolated compared 
to the others, unless O stars are hidden {\it preferentially} near the 
WN3/O3 stars.  In other words, correcting for randomly distributed but 
undetected field O stars isn't going to change the similarity between the distributions of WN3/O3 stars and RSGs, which are dominated by relatively low initial masses.
\footnote{Based on our discussion with P.~Massey after a preprint of our paper appeared, \citet{neugent17} added a footnote about the locations of WN3/O3 stars. To test the isolation that we identified, they compared the median separation between WN3/O3 stars and the nearest early WN, as compared to the median separation between WN stars and themselves. They found that the median separations were similar and concluded that there was no important difference between the origin of WN stars and WN3/O3 stars. In making this comparison, they selected a subsample of only early type WN~2-4 stars, which provide a relatively sparse background comparison sample compared to O stars. Rather than challenging the notion that WN3/O3 stars arise from relatively low initial mass, this may indicate that WN stars are heterogeneous, and that the subset of early WN~2-4 stars may also have lower initial masses than the rest. This has interesting implications and deserves more investigation, but attributing many of the early WN stars to binary evolution in a mass range just slightly above that of the WN3/O3 stars is plausible and expected from binary evolution, as discussed in the next section.}

What would be needed to make WN3/O3 stars agree with expectations of single-star evolution is to have so-far unrecognized O star clusters and associations located preferentially around the WN3/O3 sample of objects and not around the other types of evolved stars.  Since even a typical O9 main sequence star in the LMC is about a magnitude brighter at visible wavelengths than any of the WN3/O3 stars, it seems unlikely that most of the WN3/O3 stars reside amid clusters or associations of O stars that have escaped detection.

A more plausible scenario may be that selection effects make it
difficult to detect relatively faint WN3/O3 stars that reside in
crowded regions next to much brighter O-type stars.  However, one
would only expect the glare of brighter stars to hinder
the detection of fainter WN3/O3 stars if they are at small separations 
(closer than $\sim$5{\arcsec}, for example, or
1-2\,pc projected separation).  In that case, though, the population of
WN3/O3 stars would be highly bimodal, with some far on the left of
Figure~\ref{fig:cumplot}, and the rest far on the right, so we would
need to invoke two populations of WN3/O3 stars.  Barring this, it
seems that the observed isolation of WN3/O3 stars is robust, which is consistent with their similarity to the distribution of RSGs and their general avoidance of O-star clusters.

\section{EVOLUTIONARY ORIGIN}\label{evolutionary_origin}

The observed properties of WN3/O3 stars pose an intriguing puzzle for evolutionary models. Here we consider and compare various evolutionary scenarios. 

\begin{figure*}
\centering
\includegraphics[width=0.32\textwidth]{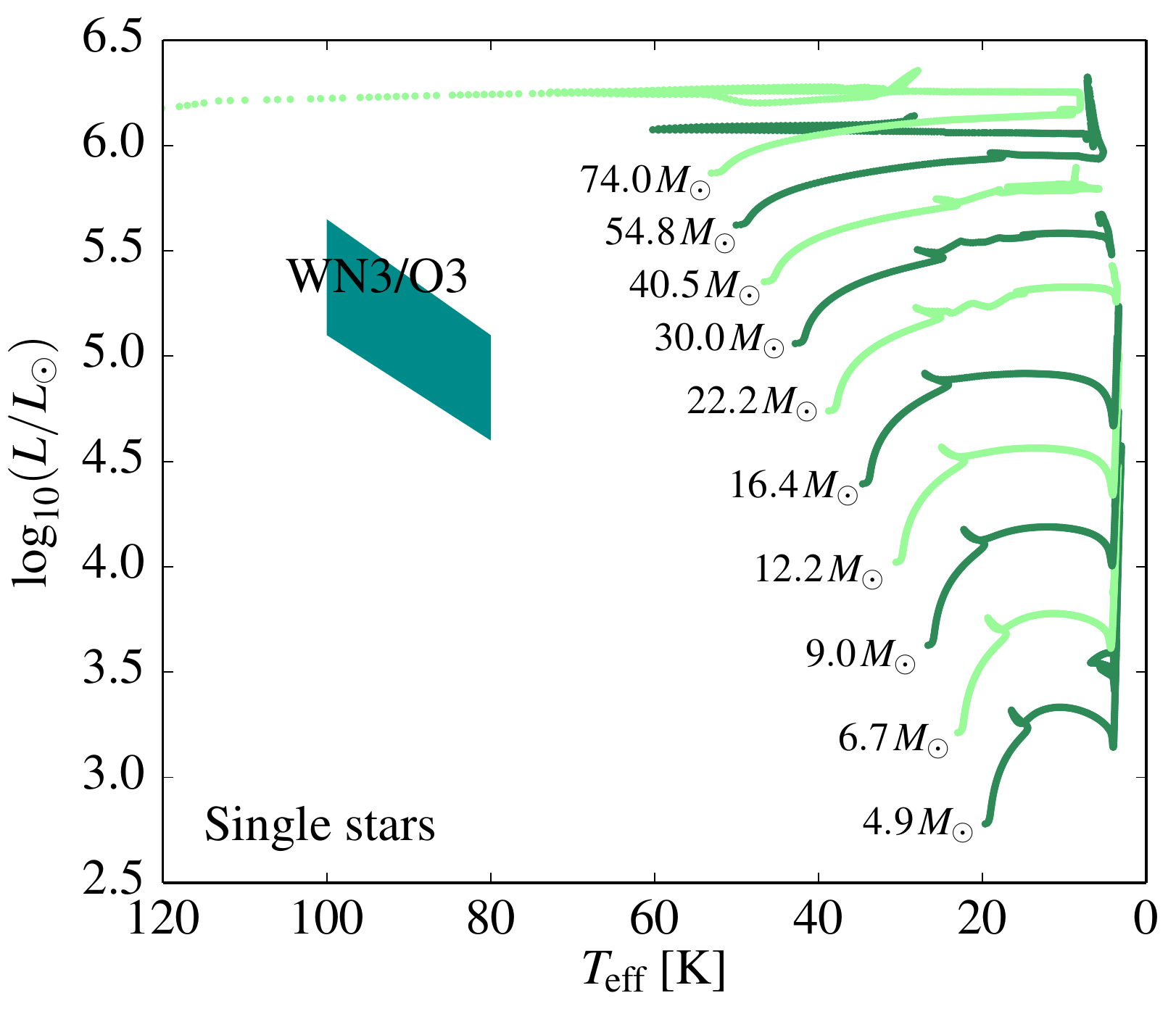}
\includegraphics[width=0.32\textwidth]{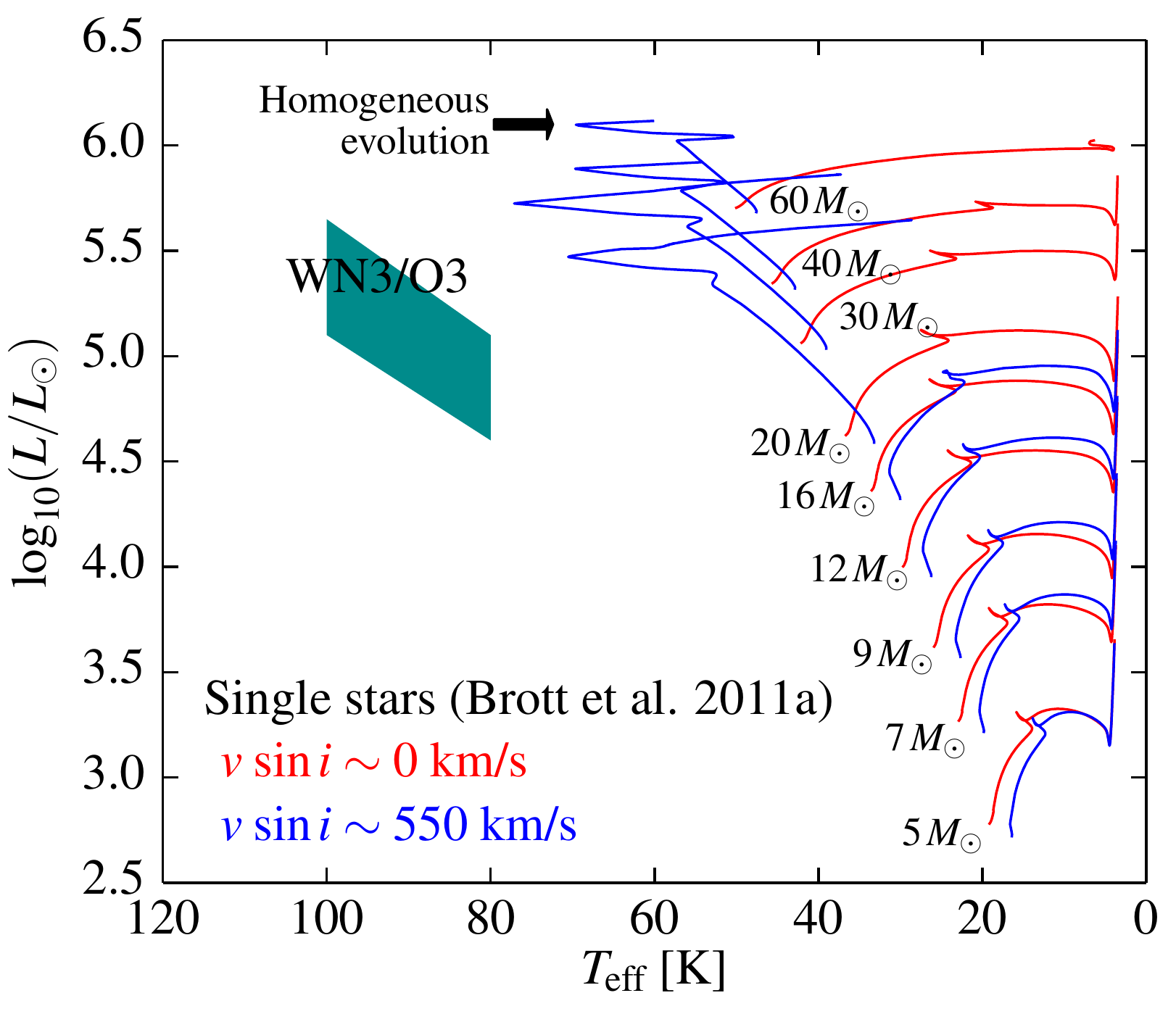}
\includegraphics[width=0.32\textwidth]{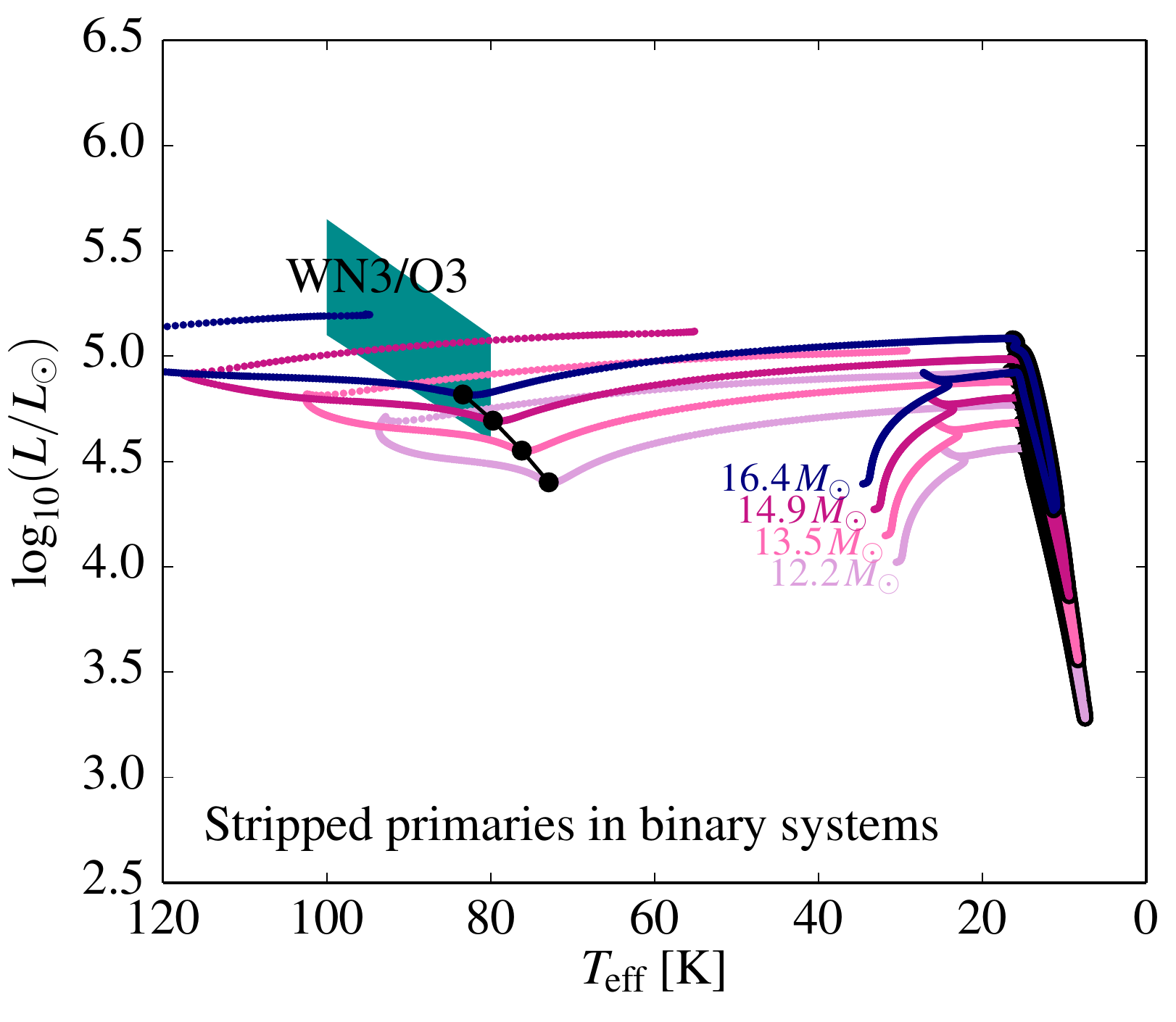}
\caption{Hertzsprung-Russell diagrams showing the approximate location of the observed WN3/O3 stars in shaded green. Left panel shows MESA evolutionary tracks for single stars (see text Section 4.1.1). Middle panel shows models of extremely rapidly rotating stars by \citet{brott11} in blue with the corresponding non-rotating tracks for comparison in red. (Note that the He burning phase of the homogeneous evolutionary tracks are not available.) Right panel shows tracks for the primary stars in binary systems 
\citep{2017A&A...608A..11G}. Black points show the location of stripped stars halfway through their He burning sequence. Labels indicate the initial masses at the onset of H burning.}
\label{fig:HRDs}
\end{figure*}

\subsection{Comparison with Evolutionary Tracks}

Figure~\ref{fig:HRDs} shows Hertzsprung-Russell (HR) diagrams, where we have marked the location of the observed WN3/O3 stars with a green  diagonal box. This locus for the WN3/O3 stars is based on the values of $T_{\rm eff}$ = 80,000 to 100,000 K and log($L$/$L_{\odot}$)=5.6 derived for the WN3/O3 star LMC170-2 by \citet{massey14}. The range in $L$ shown here is determined by the range of different observed values of $M_V$ in the full sample scaled with the same bolometric correction as LMC170-2. This location in the HR diagram is quite approximate, and further data and analysis will be needed to determine the physical parameters of the WN3/O3 stars more precisely. The implied bolometric luminosities will also be lower than shown here if it is determined that faint companions make a contribution to the $V$-band flux.  Here, we provide a preliminary comparison with evolutionary tracks.

\subsubsection{Single stars with wind mass loss}

In the left panel of Figure~\ref{fig:HRDs}, we show a sequence of evolutionary tracks for non-rotating single stars computed with the MESA evolutionary code \citep{Paxton+2011,Paxton+2013,2015ApJS..220...15P,2017arXiv171008424P}, using the same assumptions as listed in \citet{2017A&A...608A..11G}. The tracks show the evolution from the onset of hydrogen burning until the completion of central carbon burning for a metallicity of $Z=0.004$ using standard wind mass-loss rate recipes \citep{de-Jager+1988,Nieuwenhuijzen+1990, Vink+2000,Vink+2001}.

Only the most massive models shown here reach into the regime of classical WR stars, and they are too luminous for WN3/O3 stars. Allowing single star models to reach the location of the WN3/O3 stars would require  significantly higher mass-loss rates. This may be a reasonable option to consider, given the large uncertainties in our understanding of mass loss \citep[for a review see ][]{smith14}. Stars with masses between about 40 and 60\Msun may have significantly higher mass loss rates than adopted in these models if they experience LBV-like instability, which may be insensitive to metallicity \citep{so06}. However, these high-mass stars are expected to have much higher wind mass-loss rates when they reach the WR phase, as compared to the WN3/O3 stars, and LBVs are now thought to be mostly a product of binary evolution \citep{st15}. Moreover, such high-mass stars have lifetimes that are far too short to be consistent with the spatial distribution of WN3/O3 stars (Figure~\ref{fig:cumplot}); single stars with longer lifetimes at lower masses (10-20\Msun) never evolve so far to the blue. This makes it very unlikely that the WN3/O3 stars arise from traditional high-mass single stars that shed their H envelope through winds.

\subsubsection{Rotating stars evolving chemically homogeneously}

In the middle panel of Figure~\ref{fig:HRDs} we show evolutionary tracks of single stars with extremely high initial rotation rates (blue) published by \citet{brott11}. These have equatorial velocities over 500\kms \ initially, which is near their critical rotation rate. Tracks for non-rotating single stars computed with the same physical assumptions are shown for reference in red. These models adopt a metallicity $Z = 0.0047$. At the higher initial mass range, the very rapidly rotating models are assumed to experience internal mixing processes that cause them to evolve chemically homogeneously \citep{Maeder1987, Yoon+2005}. They become brighter and bluer during their main sequence evolution. After completing central He burning they are expected to contract \citep{Yoon+2006}. This is not shown in the diagram, since \citet{brott11} do not provide the complete post main sequence evolution. They will move blue-ward at nearly constant bolometric luminosity to form bright massive He stars. As above, such extremely rapidly rotating models could approach the upper bound of the area in the diagram where the brightest WN3/O3 stars reside, but it could not explain the lower end of the range of luminosity observed for WN3/O3 stars. This scenario would, however, require very extreme rotation rates and would still imply masses that are higher than we infer from their location (Section 2). This, again, makes rapidly rotating single stars a very unlikely explanation.

Instead of rapidly rotating single stars, it is worth considering whether WN3/O3 stars could be the product of rapidly rotating stars that are produced by spin-up during mass transfer in a binary system \citep{Packet1981, Cantiello+2007, de-Mink+2013}. An attractive feature of this explanation is that the rapidly rotating star can become a runaway star when its companion star explodes \citep{Hoogerwerf+2001,Blaauw1961}. This may provide a possible explanation for the extreme isolation of the brightest WN3/O3 stars, which can be tested by investigating their kinematics. If they do not have relatively large radial velocities or proper motion, then this scenario is not valid \citep[see][for a similar issue regarding LBVs]{mojgan17}. The dimmer WN3/O3 stars seem hard to reconcile with this scenario. It is also unsatisfactory that the WN3/O3 stars do not appear to show signs of rapid rotation. 

It should be noted that the theoretical concept of chemically homogeneous evolution is not without controversy. It recently gained renewed popularity in the context of gravitational wave progenitors \citep{Mandel+2016, 2016A&A...588A..50M}. However, the validity of the treatment of mixing processes in these 1D models at such high rotation rates is debated and appears to be at tension with surface abundance measurements \citep{Hunter+2009a, Brott+2011a}. Although we do not have data at present to rule out this evolutionary path, the requirement of extreme rotation rates combined with other challenges makes it seem like a rather exotic explanation in comparison to the more natural and expected scenario that we present next.

\subsubsection{Stars stripped by a binary companion}

The right panel of Figure~\ref{fig:HRDs} shows the evolution of the primary stars in binary systems that are stripped of their H envelope through Roche-lobe overflow. These models are computed with the same assumptions as detailed in \citet{2017A&A...608A..11G} for a metallicity $Z=0.004$. They are part of a larger grid that will be presented in G\"otberg et al.\ (submitted). 

For several of the evolutionary tracks for stripped star models, we find that that they spend their long-lasting core-He burning phase in the lower-luminosity end of the region of the WN3/O3 stars. These stripped stars come from initial masses between 12-16 \Msun, which is in very good agreement with the range of initial masses that we infer from their spatial distribution. This agreement makes such a scenario seem promising.

The tracks shown here only show the mass losing primary star. The companion is not shown. The observed WN3/O3 stars have their optical spectrum dominated by the He star, and do not show obvious signs of a luminous mass-gainer companion like a Be star. Below we consider two possibilities. Either the companion is a low-mass star that does not significantly contribute to the observed spectrum, or instead, the companion has already exploded. The latter requires reversal of the SN order. This has been seen in theoretical models by \citet{Pols1994a}, but only for a very limited range of parameter space. Thus, this channel is difficult to rule out, but should be rare. We consider the first possibility to be more likely, where the WN3/O3 is the remnant of a star stripped by a low-mass companion. This case implies that the companion was low mass initially and evolved through non-conservative mass transfer, preventing it from accreting a substantial fraction of the mass of the donor. This will be discussed in more detail below when we estimate rates.

\subsection{Spectral models of WN3/O3 stars}

\begin{figure}
\centering
\includegraphics[width=\hsize]{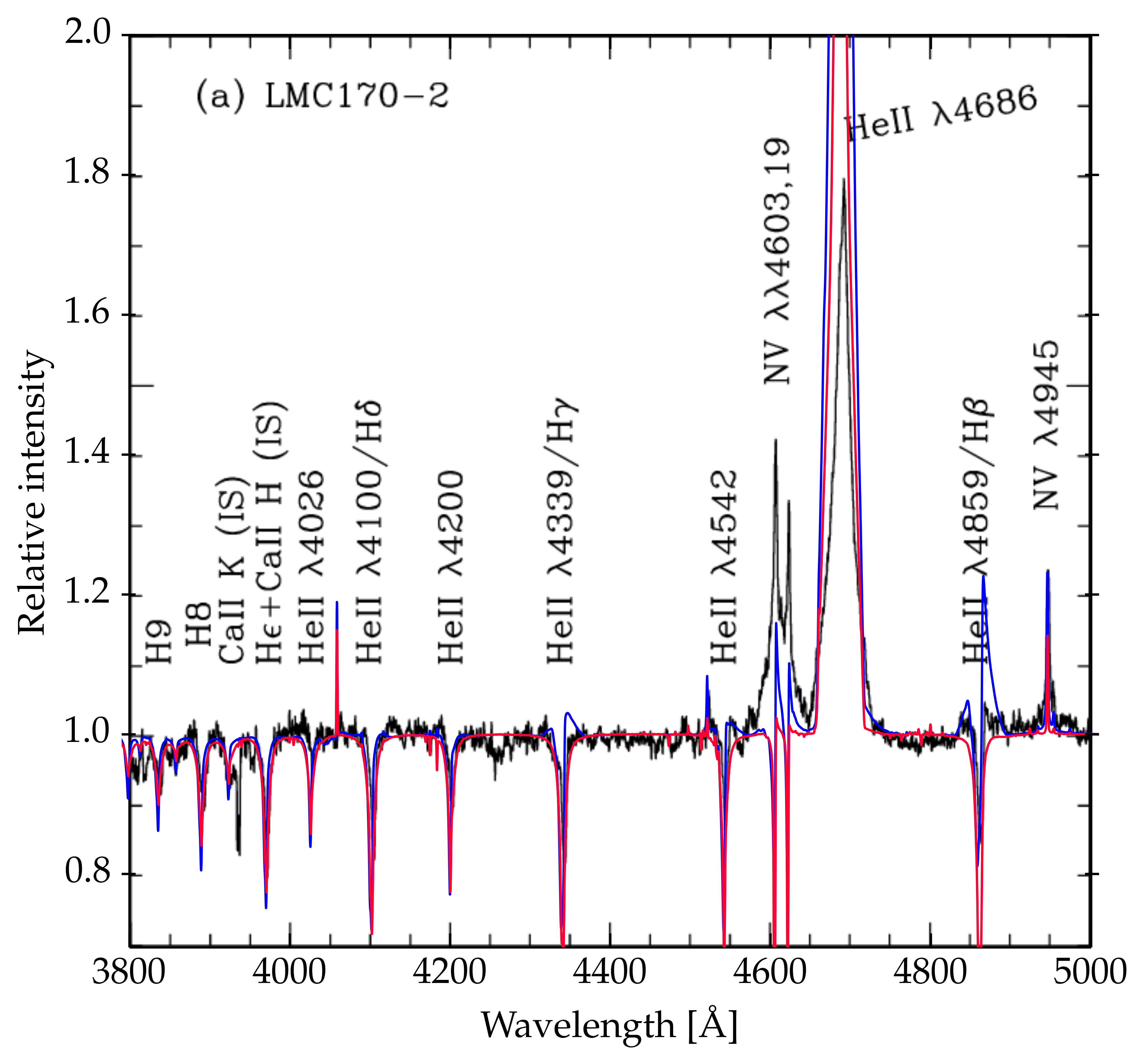}
\includegraphics[width=\hsize]{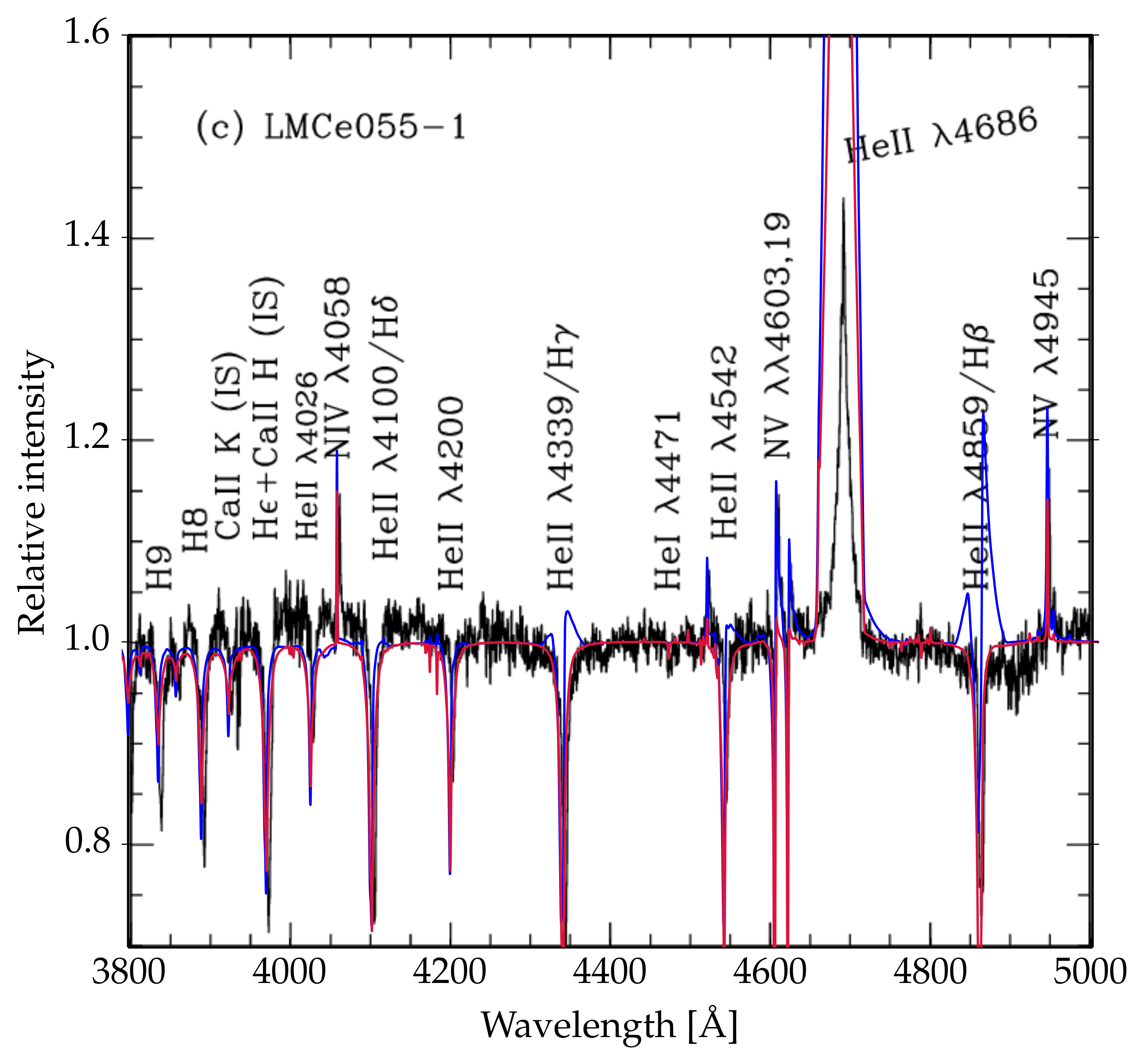}
\caption{Top: The observed WN3/O3 star LMC170-2 from \citet{massey14}
  plotted together with a stripped star model of mass 4\Msun,
  originating from an initially 12\Msun star. The red spectrum shows the nominal model, while the blue spectrum shows a model with three times higher mass loss rate. The metallicity of the model is $Z = 0.004$. Bottom: The same models as above, now plotted over the observed spectrum of the WN4/O4 star LMCe055-1, from \citet{massey17}.}
\label{fig:LMC170-2_M12}
\end{figure}

WN3/O3 stars show peculiar spectra with signatures of high $T_{\rm eff}$ and a mix of emission and absorption lines. This immediately caught our attention as resembling theoretical models recently presented by \citet{2017A&A...608A..11G}. The first spectral model computed for a stripped star was created for the quasi-WR star in HD~45166 by \citet{2008A&A...485..245G}. Other spectral models of stripped stars have been computed by \citet{2015ApJ...809..131K}, but these are for much cooler pre-SN phases, where the spectra show very different characteristics. In Figure~\ref{fig:LMC170-2_M12} we show the observed normalized spectra of two example stars (black) from \citet{massey14}. The top panel shows the WN3/O3 star LMC170-2 for which \citet{massey14} presented CMFGEN spectral modeling. The lower panel shows the somewhat cooler WN4/O4 star LMCe055-1, which is observed to be member of an eclipsing binary with a 2 day orbital period \citep{2011AcA....61..103G}.  We overplot theoretical spectra from one of the atmosphere models that we made for stars that are stripped by a binary companion. These models were computed using the CMFGEN atmosphere code \citep{Hillier+1998} for stripped stars produced with the stellar evolutionary code MESA, as detailed in \citet{2017A&A...608A..11G}, with the exception that here we use a terminal wind velocity $v_\infty = 1.9\,\times\,10^3$\kms. The models shown here are for a 4\Msun\ stripped star (i.e.\,He star) originating from a star with an initial mass of 12\Msun. The model shown in blue has a three times higher mass-loss rate compared to the nominal value shown in red. This model matches slightly better to LMC170-2, while the nominal model better matches LMCe055-1.  The wind mass-loss rate for stripped stars is uncertain as few objects have been observed. The mass-loss rate of the observed stripped star in HD~45166 agrees well with the values we assume, while theoretical models from \citet{2017A&A...607L...8V} predict lower values.

No attempt has been made to obtain a good fit to the spectrum. The model spectra were calculated previously as part of a grid of evolutionary models, and are shown here for comparison only. Interestingly, the peculiar combination of absorption and emission features that characterize the observed WN3/O3 spectra are generally reproduced very well for LMC170-2, and the correspondence is even better for LMCe055-1. The He~{\sc ii}~$\lambda$4686 flux is overestimated in the model, but the strength of the He emission features is very sensitive to the wind mass-loss rate and the terminal wind velocity, which are both uncertain quantities. In our models we use the wind mass-loss recipe by \citep{Nugis+2000} that has been extrapolated to lower $L$. For the terminal wind velocity, we use the value corresponding to 1.5 times the surface escape speed. So far, the wind mass-loss rate and terminal velocity have not been constrained by observations in this regime. It is therefore not surprising if these parameters may need some adjustment, which could be done without changing the overall properties. The atmosphere model underestimates the N~{\sc v}~$\lambda\lambda$4604,20 flux, especially in the case of LMC170-2, but it matches much better for LMCe055-1. This spectral feature is very sensitive to the wind density, ionization stage and nitrogen abundance. Reproducing this feature is a known challenge for atmosphere models \citep{2002ApJ...579..774C}. Note that the models overplotted here are taken from \citep{2017A&A...608A..11G} and were not optimized to produce a good fit. We expect that a better fit can be obtained easily by varying nitrogen abundance and wind parameters. Tuning these parameters is possible without losing the general characteristics and should lead to refined measurements of the mass-loss rate and temperature.

The atmosphere model shows emission for N~{\sc iv}~$\lambda$4058, which is observed to be in emission for LMCe055-1, but not for LMC170-2. This line is sensitive to the temperature. LMCe055-1 appears to be somewhat cooler than LMC170-2. Note that we adopted the same 80,000 - 100,000 K temperature for all WN3/O3 stars to derive their range of luminosities in Figure~\ref{fig:HRDs}. Using lower temperatures would translate to lower luminosities for some stars (since we scaled from their $M_V$), thus widening the disagreement with single-star models and rapidly rotating models even more. We expect that LMC170-2, which is on the brighter end of the distribution of visual magnitudes for WN3/O3 stars, may be better reproduced by a model of a somewhat more massive stripped star than the 12\Msun\ example shown here.

\begin{table}  \centering
  \caption{Comparison of the parameters of the spectral fit to the observed WN3/O3 star LMC170-2 as presented by \citet{massey14} with our atmosphere model for a 4\Msun stripped star originating from an originally 12\Msun~primary star in a binary system \citep{2017A&A...608A..11G}. 
  } 
  \begin{minipage}{3.3in}%
  \begin{tabular}{lcc} \hline\hline
    Parameter & Fit to LMC170-2 & Our  4 $M_{\odot}$ Model \\
    \hline
    $T_{\text{eff}}$ [kK]      & 80-100 & 73 \\
    He/H                      & $\sim$1 & 0.57 \\
    $X_{\text{N,s}} [ X_{\text{N,s, } \odot}]$     & $10 $  & $2.3 $ \\
    $\dot{M}_{\text{wind}}$ [\Msunyr] & $0.8 - 1.2 \times 10^{-6}$ & $1.6 \times 10^{-7}$ \\
    $f_{\text{vol}}$            & 0.1 & 0.5 \\
    $\log _{10} (L/L_{\odot})$   & 5.3-5.6 & 4.4 \\
    $\log _{10} (g/\text{cm\, s}^{-2})$ &... & 5.1 \\ 
    $v_{\infty}$ [\kms]          &... & $1.9 \times 10^3$\\
    $R_{\text{eff}}$ [\Rsun] &... & 1.0 
 \\    $M_V$ [mag]                  & $-2.9$ & $-0.53$ \\  \hline
\end{tabular}\label{tab:prop}
\medskip 
{\small \textbf{Notes.} We use the following notation for the stellar properties: $T_{\text{eff}}$ for effective temperature, He/H for relative number fraction, $X_{\text{N,s}}$ for surface nitrogen mass fraction displayed in comparison to the solar value, $\dot{M}_{\text{wind}}$ for wind mass-loss rate, $f_{\text{vol}}$ for volume filling factor in wind clumping, $L$ for luminosity, $\log _{10} g$ for surface gravity, $v_{\infty}$ for terminal wind velocity, $R_{\text{eff}}$ for effective radius.}
\end{minipage}
\end{table}

In Table~\ref{tab:prop} we compare the parameters for our atmosphere model to those presented by \cite{massey14}, derived from their CMFGEN fit to LMC170-2.  The mass-loss rate, $T_{\rm eff}$, and luminosity are lower in our model.  
Our model also underestimates the brightness of LMC170-2 in the $V$-band. This may indicate that this star is somewhat more massive than our fiducial model, as noted above. Alternatively, it might signify a contribution of visual-wavelength flux by a low-mass main-sequence companion star. In binary models, the secondary star is likely to be a rapid rotator as a result of mass accretion. Such a faint companion with broad spectral features may be difficult to recognize in the optical spectrum. 

Despite the fact that the models were not fit to the data, but rather that their physical parameters were predetermined by stellar evolutionary models, we find that they do a remarkable job of achieving a mixture of the absorption and emission that is qualitatively similar to the observed spectra. There are of course parameters in the model that can be adjusted to give a better fit to individual line strengths, such as the mass-loss rate, wind clumping, and temperature. Refining these parameters was not our goal, but this could be achieved in future studies when higher quality data are available.

\subsection{Rate estimates}

To explain WN3/O3 stars as stripped He stars we need them to either have a faint companion, or to be the former member of a disrupted binary system, as discussed above. The second possibility requires the secondary to explode before the primary, which should be very rare according to current standard simulations \citep{Pols1994a, Claeys+2011}, although this does depend on the treatment of convection and rejuvenation processes \citep[e.g.][]{Braun+1995, Dray+2007}. The case of a faint low-mass companion on non-conservative mass transfer may also be somewhat rare, but more likely than a reversal of the SN order.     

A lower mass companion should be a less evolved star that still resides on the main sequence. Comparing the brightness of low-mass main-sequence star models \citep{Kurucz1992} to the model of stripped stars at optical wavelengths, we find that the spectral model of a stripped star presented above is brighter at optical wavelengths than main-sequence stars with a late B spectral type (or fainter), suggesting the companion stars to have initial masses $M_2 \lesssim 4$\Msun. Systems with such extreme mass ratios and relatively long periods \citep[more than about 300 days, ][]{Schneider+2015}, are thought to evolve through highly non-conservative mass transfer. This means that the companion will not gain a significant amount of mass, so that it remains faint even after mass transfer. It may, however, gain significant angular momentum to become a rapid rotator, as noted above.

Below, we estimate the relative number of stars that are stripped by a low-mass companion, based on the rough constraints on the initial period and mass ratio stated above. This allows us to provide a very rough estimate for the relative fraction of stars with initial mass of $\sim$12 \Msun \ that become a stripped He star, with respect to 12 \Msun \ stars that spend their He burning lifetime as RSGs.  

\begin{enumerate}

\item  Massive stars are typically found with close companions. We assume that half of all 12\Msun~stars are the primary of a binary system, $f_{\text{bin}} \approx 0.5$ \citep{Duchene+2013,2016arXiv160605347M}.

\item For an initially 12\Msun \ primary star, we estimate that the faint companion probably is less massive than about 4\Msun. We assume a uniform distribution of mass ratios, which is consistent with observations \citep{Kobulnicky+2014}. The fraction of systems with mass ratios in the appropriate range is thus $f_q \approx 0.25$. 

\item We require initially wide binary systems, larger than about 300 days, to ensure non-conservative mass transfer and avoid a merger between the two stars \citep{Schneider+2015}. (Note that 300 days is not necessarily the final period.)  We assume the initial orbital periods to be distributed according to \"Opiks law, which is a uniform distribution in the log of the period \citep{opik24}, between about 3 and 3000 days \citep{Kobulnicky+2014, Dunstall+2015}. The fraction of systems wider than 300 days is thus $f_P \approx 0.33$. 
  
\item Stripped stars spend approximately 10\% of their lifetime in the stripped star phase. This time is set by the duration of the He burning phase. For stars that are not stripped, we assume they spend their lifetime as RSGs, with a similar duration. Since the lifetimes for each are comparable we do not need to explicitly account for this in our rate estimate.

\end{enumerate}

The above considerations imply the ratio of observable stripped stars to RSGs resulting from stars with the same initial mass of approximately 12\Msun, $f_{\rm strip} = N_{\rm strip} / N_{\rm RSG}$, is a few percent, 

$$f_{\rm strip} \sim 0.04 \times \left(\frac{f_{\text{bin}}}{0.5} \right) \left(  \frac{f_q}{0.25} \right) \left(  \frac{f_P}{0.33} \right). $$

Comparing this directly to the currently observed number of WN3/O3 stars, $N_{\rm strip, obs} = 10$, is tricky given the large uncertainties, biases that are not yet well understood, and incompleteness.
\citet{2012ApJ...749..177N} find of order 500 RSGs in the LMC, although these RSGs may span a wider mass range than the stripped binaries that we consider. Also the completeness fraction, $f_{\rm comp}$ of the survey of \citet{massey14} is uncertain for our purposes. In terms of finding stripped stars, \citeauthor{massey14}'s search is probably not limited so much by the $V$ magnitude of the stars, because they noted that the detection limit was significantly fainter than the WN3/O3 stars. Fainter stripped stars should exist, but they will also have weaker winds, and hence, weaker emission lines. The weak emission lines rather than the fainter continuum may cause them to be missed in a narrow-band imaging survey.  Of course, \citeauthor{massey14}'s survey is not yet finished; significant portions of the LMC have yet to be surveyed, so it is possible that more WN3/O3 stars will be discovered in the LMC. 

Nevertheless, adopting very rough but plausible fiducial number estimates for the comparison RSGs, the fraction of stars in this mass range that are stripped primaries, and some completeness factor can yield a number of stripped stars that matches the sample of WN3/O3:   
$$
N_{\rm strip, obs}  \sim 10 \times
\left(\frac{N_{\rm RSG}}{500} \right)
\left(\frac{f_{\rm strip}}{0.04} \right)
\left(\frac{f_{\rm comp}}{0.5} \right)
$$
We emphasize that is an extremely crude estimate, which should only be considered as an order of magnitude assessment. We conclude from this that the number of WN3/O3 stars found so far agrees reasonably well with the number of stripped primaries with low-mass companions that one expects from a population.

Interestingly, we expect a large majority of stripped stars to be difficult to see because they have a more massive mass-gainer companion that is too bright. The detection of these stripped stars will be discussed in a future paper, but in some of these cases we predict FUV-excess detection to be a promising technique.

\subsection{Detection of companions}

A strong clue that binarity may play a role in the formation of WN3/O3 stars comes from the system LMCe055-1, which is an observed eclipsing binary \citep{2011AcA....61..103G,massey17}, with a period of only $\sim$2.2\,days. Such a short orbital period is expected in the case of common envelope ejection. Obtaining the detailed system parameters for this system will be of great value to constrain models of mass transfer.  

In general, a low-mass companion may be easily overlooked. It would not dominate the total flux at near-UV or visual wavelengths (by design). Given that its mass is low, its gravitational pull on the WN3/O3 star is also small. For wider orbits, radial velocities would have a low amplitude and would only be detectable with multi-epoch, high-resolution spectroscopy. Even then, the system requires a favorable inclination. Finally, there is still the option that these stripped He stars come from disrupted systems with no remaining companion. Such systems may stand out by a peculiar radial velocity, but for systems that are wide at the moment of disruption of the supernova, the radial velocity may be much less than the canonical threshold of 30\kms for runaway stars.

\citet{massey14} note that LMC143-1 and  LMC173-1 are classified as binaries containing WN3 + late O-type binaries (they are also brighter, and we do not consider them among the ten WN3/O3 stars discussed here; see Table~\ref{tab:list}). They mention an absence of evidence for binarity for the other stars, although this is discussed mostly in the context of binarity as an inadequate explanation for the composite emission/absorption nature of the observed spectra. Low-mass companions are much harder to rule out. For most of the objects in the sample, \citet{massey14} present only a single epoch of spectroscopy, which makes it difficult to detect radial velocity variations. Moreover, the spectra have only low/moderate resolution, with a resolving power $R \sim 4100$ \citep{massey14,massey15,massey17}. With these parameters, only large variations would be detectable for the short-period systems that have multiple epochs, even with an optimal cadence. At this resolving power, low-amplitude radial velocity variations from a low-mass companion might have been overlooked.
As an example, for a 4\Msun \, stripped star and 2 \Msun \, companion, the resolving power requires an orbit shorter than 10\,days in order to detect the companion. This may apply to systems that have evolved through common envelope ejection, but systems that formed through stable non-conservative mass transfer may well have orbits that are substantially larger than this.  

We conclude that observational constraints so far do not preclude a binary origin for WN3/O3 stars. Dedicated multi-epoch spectroscopy at high resolution will be required to further investigate the binary nature of these objects.

\subsection{Summary and testable predictions}

We considered the hypotheses that WN3/O3 stars result from single stars, rapidly rotating stars that evolve homogeneously, or stripped stars in binary systems. We find that the last possibility most naturally explains (1) the remote location of the WN3/O3 stars and the implications for their progenitor masses, (2) the peculiar spectra showing a mix of emission and absorption lines, and (3) the fact that they are under luminous compared to normal WR stars. We further find that this scenario (4) predicts rates that are roughly the right order of magnitude, and (5) is consistent with the current constraints on detections of companions for WN3/O3 stars presented by \citet{massey17}. A critical next step is to test our hypothesis and verify whether WN3/O3 stars are indeed binary systems containing a stripped star together with a faint companion, or whether they exhibit any peculiar spatial velocities indicating that they come from disrupted binary system.  If the stripped stars still have faint companions, the majority of the sample should have detectable radial velocity variations. This will require high-resolution spectra, since the variations may be small in many cases, as noted above. 

For those systems that may have formed from binary systems where the secondary exploded first, their isolated environments require that they are slow runaways, and so they should exhibit peculiar space velocities. However, a complication is that reversal of the SN order requires conservative mass transfer and implies substantial widening of the system before explosion. The spatial velocities of such systems may not necessarily be high enough to be classified as runaway stars ($> 30$\kms) and may be difficult to distinguish as peculiar from the surrounding stellar population. Future proper motions with high precision may be a crucial test in this regard.

Detection of eclipses or ellipsoidal variations will be especially useful as they will provide tight constraints on the system parameters. LMCe055-1 is therefore an extremely interesting system for further study.   However, eclipses are only expected for the closest systems whose orbits are favorably aligned.

\section{SUPERNOVA PROGENITORS}

WN3/O3 stars are less luminous and have lower mass-loss rates than 
classical WN stars.  In this paper, we have shown that they also have 
much more isolated environments relative to O-type stars in the LMC.  
This isolation requires that they have longer
lifetimes and substantially lower initial masses than classical WN
stars, comparable to normal RSGs with initial masses of 10-15\Msun.  
This immediately points to binary evolution as a way to
produce these H-poor stars in this mass range.  In considering a wide
range of binary scenarios, we find that a plausible evolutionary
channel to explain WN3/O3 stars is that stripping of the primary
star's H envelope occurs in a system where the secondary star is a
relatively low-mass main sequence star ($<$4\Msun), and where
the separation is initially wide, such that mass transfer is non-conservative.
In this case, the companion is substantially fainter than the WN3/O3
star primary and could be hidden in the spectrum.  This is expected to
be a minority of all cases of binary-stripped stars, but it is a
majority of cases without a bright companion where the stripped star 
dominates the visual spectrum. We argued that the
expected number of such stars in the LMC, while uncertain, is
plausibly in agreement with the observed WN3/O3 stars.

In addition to matching the properties of WN3/O3 stars, the lower-mass
stripped progenitors in binaries that we describe also do a good job
of fitting expectations for the progenitors of most stripped-envelope
SNe, especially those of Type~Ib.  Our example cases had initial
masses of roughly 12$-$16\Msun \ and final masses as stripped He stars of only 
4$-$5\Msun.  If these stars were to explode, and if $\sim$1.5\Msun \ 
goes into the neutron star, then they would have He-rich
ejecta masses of roughly 2-2.5\Msun.  This agrees well with the
average value of derived ejecta mass for SNe~Ib, which is
$\sim$2~$M_{\odot}$ \citep{drout11,lyman16}.  All the WN3/O3 stars in
the LMC have absolute magnitudes that are fainter (roughly $-$2 to
$-$3 mag) than published upper limits ($-$5 mag) for non-detections of
Type~Ib progenitors in pre-SN imaging \citep{smartt09}, so they are
consistent with available constraints from direct imaging.

As noted in the Introduction, many lines of evidence now suggest that
a large fraction of stripped-envelope SNe arise from relatively
low-mass progenitors (initially 10-20\Msun) in binary systems
rather than from more massive progenitors ($>$35\Msun) that have
lost their H in winds to become classical WR stars before exploding.
The lack of any direct
detection (thus far) of a massive WR star as a stripped-envelope SN
progenitor is consistent with a dominant binary scenario,
although this is not necessarily conclusive since hot progenitors may
be hard to detect in available progenitor imaging. The only reported detection of a progenitor to a stripped-envelope Type Ib explosion is the case of iPTF13bvn, where its progenitor is thought to have been a lower-mass helium giant \citep{2016MNRAS.461L.117E,2016ApJ...825L..22F}.

SNe~IIb are also stripped-envelope SNe, but in these cases a small
mass of residual H (about 0.1\Msun \ or less) remains in the
star's envelope at death.  Direct detections of SNe~IIb progenitors
and some of their potential surviving companions point to modest
initial masses of 10-20\Msun and binary origins for this
sub-class of stripped stars as well
\citep{aldering94,maund04,maund11,svd11,svd14,fox14}.

Interestingly, two recent studies have searched for surviving
companions of nearby SNe~Ic (SN~1994I and SN~2002ap) and both find
that any companion is less massive than $\sim$10\Msun \ 
\citep{svd16,zapartas17}. These systems require stripping in a binary
with a low-mass companion that does not gain much mass,
similar to the WN3/O3 star scenario that we discuss above.

The fact that WN3/O3 stars have a statistical separation from O-type
stars resembling that of relatively low-mass RSGs is also interesting
in comparison to the observed statistics of SN environments.  While
the locations of SNe~Ic and SNe~Ic-BL within their host galaxies tend
to favor relatively young ages and high initial masses, the
environments of SNe~Ib are distinct and appear to be more like those
of SNe~II-P that arise mostly from older stars
\citep{aj08,anderson12}.

Another SN type for which the WN3/O3 stars are potential progenitors
is the class of Type~Ibn explosions.  These are stripped-envelope SNe
that lack H, but have strong narrow He lines
\citep{pastorello08,hoss17}. The narrow He lines are thought to be
indicative of shock interaction with circumstellar material (CSM),
requiring pre-SN eruptive mass loss akin to Type~IIn supernovae
\citep{smith14}, but occurring in systems with a stripped-envelope
progenitor.\footnote{Note that WN3/O3 stars do not yet have the dense CSM required to yield a SN~Ibn if they were to explode in their
  current state.  This material is ejected in precursor eruptions a
  few years before the SN.  Our point here is only that SNe~Ibn may have
  moderately massive (initially 10-20\Msun) and sometimes isolated
  stripped envelope progenitors, similar to WN3/O3 stars.}
Information about the environments of SNe~Ibn is still admittedly
somewhat anecdotal, since these are relatively rare and not many are
known yet.  However, it does appear that a subset prefers relatively
faint, blue hosts and relatively remote locations within host
galaxies.  In particular, the two best-studied SNe~Ibn so far,
SN~2006jc \citep{pastorello07,foley07} and SN~2015G
\citep{hoss17,shivvers17} occurred in very sparse outer regions of
their hosts.  This is reminiscent of some extremely isolated SNe~IIn,
such as SN~2009ip \citep{smith+16}. One SN~Ibn was even seen to occur
in an early-type galaxy \citep{sanders13}.  Thus, it is not clear that
SN~Ibn environments necessarily point to massive WR-like progenitors.  A quantitative
analysis of SN~Ibn host environments has not yet been conducted,
however, so this link is tentative.

Thus, it is intriguing that the very isolated environments of WN3/O3
stars (at least compared to fairly massive O-type stars) implies that
they fit the bill for binary-stripped progenitors from initially
lower-mass stars than classical WR stars.  They could be good examples
of stripped-envelope SN progenitors where the companion star is too
faint to be detected after a SN fades.  We argue that this is most
commonly because the companion is a faint, low-mass star where the
system has experienced non-conservative mass transfer.  

There will be more common cases where the secondary gains a significant amount of mass or had a relatively high initial mass, in which case the companion may be so bright at optical wavelengths that it may be difficult to see the hot stripped star.  Searches for nearby counterparts of these hot stripped stars hidden by bright companions are ongoing.  So far, only a handful candidates are known: $\phi$~Per, FY~CMa, 59~Cyg, 60~Cyg, HR~2142, and HD~45166 \citep[\rev{see}][respectively]{1998ApJ...493..440G, 2008ApJ...686.1280P, peters13, 2017ApJ...843...60W, 2016ApJ...828...47P, 2005A&A...444..895S, 2008A&A...485..245G}. In all these systems, the candidate hot He star is extremely difficult to study because it orbits a cooler and brighter B or Be star.
Thus, the fact that WN3/O3 stars seem to
outshine their faint, low-mass companion provides an extremely valuable
opportunity for detailed investigations of the physical properties of likely
SN~Ib progenitor stars.  

Since they are relatively faint and hot, perhaps we are still missing many of 
the WN3/O3 stars in nearby environments.  It would be interesting to try and 
detect more examples in the Milky Way and SMC.  If they do not exist in the 
Milky Way (or if they are pushed to even lower initial masses), it may be 
because stronger winds at higher metallicity lead to emission line spectra that
hide the O3~V-type absorption signatures, or alternatively, because of different 
physics of mass transfer due to more effective cooling at higher metallicity.
Similarly, in the SMC, a strong metallicity dependence \citep{2017IAUS..329..171H} might 
push the WN3/O3 phenomenon to higher luminosity.  Indeed, about half the known 
WR stars in the SMC have spectra that closely resemble those of the LMC WN3/O3 
stars.   In any case, it will be interesting to increase the sample size in 
order to explore their statistical properties further, and to identify them in 
other host galaxies to explore their metallicity dependence.

\section{SUMMARY}

We investigate the nature of WN3/O3 stars in the LMC reported by Massey et 
al.\ in three recent papers \citep{massey14,massey15,massey17}.  Using an 
analysis of their environments similar to the study of LBVs by \citet{st15}, 
we find that their observed distribution of separations from O-type stars on
the sky shows that they are much more isolated than other types of WR
stars.  Their projected separations from O-type stars on the sky is
similar to that of normal RSGs in the LMC with initial masses of only 
10-15\Msun.  This requires that the observed WN3/O3 stars have similar 
lifetimes and come from a similar initial mass range as typical RSGs.  This 
class is therefore distinct from the classical WR stars (WN and WC) thought 
to arise from higher initial mass (i.e.\ roughly $>$ 32\Msun\ for WN and 
$>$50\Msun\ for WO/WC \citealt{groh14}).  In this lower mass range, 
radiation-driven stellar winds are too weak for the WN3/O3 stars to have shed 
their H envelopes via winds.  Their relative isolation and longer lifetimes 
therefore suggest that they have been stripped of the H envelopes through binary evolution.

Considering a range of binary evolution models, we propose a plausible
scenario for the evolutionary origin of WN3/O3 stars, where they are
the stripped primary stars in binary systems with a relatively low-mass 
secondary and relatively wide initial orbital separation.  This leads to 
highly non-conservative mass transfer, so that after the mass transfer phase
has ended, the secondary has accreted little mass and remains faint,
whereas the more luminous primary has been stripped of its H envelope
and appears as a WN3/O3 star.  The secondary contributes a small
fraction of the total visible light, and radial velocity variations due to
orbital motion are likely to be small in many cases, allowing them to easily 
have escaped detection so far.  We suggest that monitoring of these systems 
with high spectral resolution may reveal the presence of faint companions, 
and can provide a direct test of our proposed origin for WN3/O3 stars.

Finally, we note that WN3/O3 stars are excellent candidates for SN~Ib
progenitors.  Because of their low-mass and (presumed) faint
secondary, they are more readily observable than stripped stars with a
higher-mass companion that has gained mass.  They therefore offer a
way to constrain the physical properties of likely progenitors for Type Ib
and possibly also Type Ibn explosions, both of which seem to favor
more isolated environments than SNe~Ic and SNe~Ic-BL.  It would be
good to search for WN3/O3 stars in the Milky Way, or if they cannot be
detected, to understand why they may not exist at the same luminosities
at higher metallicity.

\section*{Acknowledgements}

\scriptsize 

The authors wish to acknowledge Jose Groh, Norbert Langer, Hugues Sana, Manos Zapartas, Mathieu Renzo, Alex de Koter, and Max Moe for numerous stimulating discussions.  Support for NS was provided by NSF awards AST-1312221 and AST-1515559, by a  Scialog grant from the Research Corporation for Science Advancement, and by the National Aeronautics and Space Administration (NASA)
through HST grant AR-14316 from the Space Telescope Science Institute,
which is operated by AURA, Inc., under NASA contract NAS5-26555. 
SdM has received funding  under the European Union’s Horizon 2020 research and innovation programme from the European under the Marie Skłodowska-Curie (Grant Agreement No. 661502) and the European Research Council (ERC) (Grant agreement No. 715063). The authors further acknowledge hospitality of the Kavli Institute for 
Theoretical physics, Santa Barbara, Ca. Their stay was supported by the 
National Science Foundation under Grant No. NSF PHY11-25915.  This research has made use of the SIMBAD data base, operated at CDS,
Strasbourg, France.

\bibliographystyle{mnras.bst}
\bibliography{my_bib,nathan_new_bib,references_bin}

\begin{thebibliography}{}
\makeatletter
\relax
\def\mn@urlcharsother{\let\do\@makeother \do\$\do\&\do\#\do\^\do\_\do\%\do\~}
\def\mn@doi{\begingroup\mn@urlcharsother \@ifnextchar [ {\mn@doi@}
  {\mn@doi@[]}}
\def\mn@doi@[#1]#2{\def\@tempa{#1}\ifx\@tempa\@empty \href
  {http://dx.doi.org/#2} {doi:#2}\else \href {http://dx.doi.org/#2} {#1}\fi
  \endgroup}
\def\mn@eprint#1#2{\mn@eprint@#1:#2::\@nil}
\def\mn@eprint@arXiv#1{\href {http://arxiv.org/abs/#1} {{\tt arXiv:#1}}}
\def\mn@eprint@dblp#1{\href {http://dblp.uni-trier.de/rec/bibtex/#1.xml}
  {dblp:#1}}
\def\mn@eprint@#1:#2:#3:#4\@nil{\def\@tempa {#1}\def\@tempb {#2}\def\@tempc
  {#3}\ifx \@tempc \@empty \let \@tempc \@tempb \let \@tempb \@tempa \fi \ifx
  \@tempb \@empty \def\@tempb {arXiv}\fi \@ifundefined
  {mn@eprint@\@tempb}{\@tempb:\@tempc}{\expandafter \expandafter \csname
  mn@eprint@\@tempb\endcsname \expandafter{\@tempc}}}

\bibitem[\protect\citeauthoryear{{Aghakhanloo}, {Murphy}, {Smith}  \& {Hlo{\v
  z}ek}}{{Aghakhanloo} et~al.}{2017}]{mojgan17}
{Aghakhanloo} M.,  {Murphy} J.~W.,  {Smith} N.,   {Hlo{\v z}ek} R.,  2017,
  \mn@doi [\mnras] {10.1093/mnras/stx2050}, \href
  {http://adsabs.harvard.edu/abs/2017MNRAS.472..591A} {472, 591}

\bibitem[\protect\citeauthoryear{{Aldering}, {Humphreys}  \&
  {Richmond}}{{Aldering} et~al.}{1994}]{aldering94}
{Aldering} G.,  {Humphreys} R.~M.,   {Richmond} M.,  1994, \mn@doi [\aj]
  {10.1086/116886}, \href {http://adsabs.harvard.edu/abs/1994AJ....107..662A}
  {107, 662}

\bibitem[\protect\citeauthoryear{{Anderson} \& {James}}{{Anderson} \&
  {James}}{2008}]{aj08}
{Anderson} J.~P.,  {James} P.~A.,  2008, \mn@doi [\mnras]
  {10.1111/j.1365-2966.2008.13843.x}, \href
  {http://adsabs.harvard.edu/abs/2008MNRAS.390.1527A} {390, 1527}

\bibitem[\protect\citeauthoryear{{Anderson}, {Habergham}, {James}  \&
  {Hamuy}}{{Anderson} et~al.}{2012}]{anderson12}
{Anderson} J.~P.,  {Habergham} S.~M.,  {James} P.~A.,   {Hamuy} M.,  2012,
  \mn@doi [\mnras] {10.1111/j.1365-2966.2012.21324.x}, \href
  {http://adsabs.harvard.edu/abs/2012MNRAS.424.1372A} {424, 1372}

\bibitem[\protect\citeauthoryear{{Blaauw}}{{Blaauw}}{1961}]{Blaauw1961}
{Blaauw} A.,  1961, \bain, \href
  {http://adsabs.harvard.edu/abs/1961BAN....15..265B} {15, 265}

\bibitem[\protect\citeauthoryear{{Braun} \& {Langer}}{{Braun} \&
  {Langer}}{1995}]{Braun+1995}
{Braun} H.,  {Langer} N.,  1995, \aap, \href
  {http://adsabs.harvard.edu/abs/1995A%26A...297..483B} {297, 483}

\bibitem[\protect\citeauthoryear{{Brott} et~al.,}{{Brott}
  et~al.}{2011a}]{brott11}
{Brott} I.,  et~al., 2011a, \mn@doi [\aap] {10.1051/0004-6361/201016113}, \href
  {http://adsabs.harvard.edu/abs/2011A%26A...530A.115B} {530, A115}

\bibitem[\protect\citeauthoryear{{Brott} et~al.,}{{Brott}
  et~al.}{2011b}]{Brott+2011a}
{Brott} I.,  et~al., 2011b, \mn@doi [\aap] {10.1051/0004-6361/201016114}, \href
  {http://adsabs.harvard.edu/abs/2011A%26A...530A.116B} {530, A116}

\bibitem[\protect\citeauthoryear{{Cantiello}, {Yoon}, {Langer}  \&
  {Livio}}{{Cantiello} et~al.}{2007}]{Cantiello+2007}
{Cantiello} M.,  {Yoon} S.-C.,  {Langer} N.,   {Livio} M.,  2007, \mn@doi
  [\aap] {10.1051/0004-6361:20077115}, \href
  {http://adsabs.harvard.edu/abs/2007A%26A...465L..29C} {465, L29}

\bibitem[\protect\citeauthoryear{{Claeys}, {de Mink}, {Pols}, {Eldridge}  \&
  {Baes}}{{Claeys} et~al.}{2011}]{Claeys+2011}
{Claeys} J.~S.~W.,  {de Mink} S.~E.,  {Pols} O.~R.,  {Eldridge} J.~J.,   {Baes}
  M.,  2011, \mn@doi [\aap] {10.1051/0004-6361/201015410}, \href
  {http://adsabs.harvard.edu/abs/2011A%26A...528A.131C} {528, A131}

\bibitem[\protect\citeauthoryear{{Conti}}{{Conti}}{1975}]{conti76}
{Conti} P.~S.,  1975, Memoires of the Societe Royale des Sciences de Liege,
  \href {http://adsabs.harvard.edu/abs/1975MSRSL...9..193C} {9, 193}

\bibitem[\protect\citeauthoryear{{Crowther}}{{Crowther}}{2007}]{crowther07}
{Crowther} P.~A.,  2007, \mn@doi [\araa]
  {10.1146/annurev.astro.45.051806.110615}, \href
  {http://adsabs.harvard.edu/abs/2007ARA%26A..45..177C} {45, 177}

\bibitem[\protect\citeauthoryear{{Crowther}, {Hillier}, {Evans}, {Fullerton},
  {De Marco}  \& {Willis}}{{Crowther} et~al.}{2002}]{2002ApJ...579..774C}
{Crowther} P.~A.,  {Hillier} D.~J.,  {Evans} C.~J.,  {Fullerton} A.~W.,  {De
  Marco} O.,   {Willis} A.~J.,  2002, \mn@doi [\apj] {10.1086/342877}, \href
  {http://adsabs.harvard.edu/abs/2002ApJ...579..774C} {579, 774}

\bibitem[\protect\citeauthoryear{{Dessart}, {Hillier}, {Livne}, {Yoon},
  {Woosley}, {Waldman}  \& {Langer}}{{Dessart} et~al.}{2011}]{dessart11}
{Dessart} L.,  {Hillier} D.~J.,  {Livne} E.,  {Yoon} S.-C.,  {Woosley} S.,
  {Waldman} R.,   {Langer} N.,  2011, \mn@doi [\mnras]
  {10.1111/j.1365-2966.2011.18598.x}, \href
  {http://adsabs.harvard.edu/abs/2011MNRAS.414.2985D} {414, 2985}

\bibitem[\protect\citeauthoryear{{Dray} \& {Tout}}{{Dray} \&
  {Tout}}{2007}]{Dray+2007}
{Dray} L.~M.,  {Tout} C.~A.,  2007, \mn@doi [\mnras]
  {10.1111/j.1365-2966.2007.11431.x}, \href
  {http://adsabs.harvard.edu/abs/2007MNRAS.376...61D} {376, 61}

\bibitem[\protect\citeauthoryear{{Drout} et~al.,}{{Drout}
  et~al.}{2011}]{drout11}
{Drout} M.~R.,  et~al., 2011, \mn@doi [\apj] {10.1088/0004-637X/741/2/97},
  \href {http://adsabs.harvard.edu/abs/2011ApJ...741...97D} {741, 97}

\bibitem[\protect\citeauthoryear{{Duch{\^e}ne} \& {Kraus}}{{Duch{\^e}ne} \&
  {Kraus}}{2013}]{Duchene+2013}
{Duch{\^e}ne} G.,  {Kraus} A.,  2013, \mn@doi [\araa]
  {10.1146/annurev-astro-081710-102602}, \href
  {http://adsabs.harvard.edu/abs/2013ARA%26A..51..269D} {51, 269}

\bibitem[\protect\citeauthoryear{{Dunstall} et~al.,}{{Dunstall}
  et~al.}{2015}]{Dunstall+2015}
{Dunstall} P.~R.,  et~al., 2015, \mn@doi [\aap] {10.1051/0004-6361/201526192},
  \href {http://adsabs.harvard.edu/abs/2015A%26A...580A..93D} {580, A93}

\bibitem[\protect\citeauthoryear{{Eldridge} \& {Maund}}{{Eldridge} \&
  {Maund}}{2016}]{2016MNRAS.461L.117E}
{Eldridge} J.~J.,  {Maund} J.~R.,  2016, \mn@doi [\mnras]
  {10.1093/mnrasl/slw099}, \href
  {http://adsabs.harvard.edu/abs/2016MNRAS.461L.117E} {461, L117}

\bibitem[\protect\citeauthoryear{{Folatelli} et~al.,}{{Folatelli}
  et~al.}{2016}]{2016ApJ...825L..22F}
{Folatelli} G.,  et~al., 2016, \mn@doi [\apjl] {10.3847/2041-8205/825/2/L22},
  \href {http://adsabs.harvard.edu/abs/2016ApJ...825L..22F} {825, L22}

\bibitem[\protect\citeauthoryear{{Foley}, {Smith}, {Ganeshalingam}, {Li},
  {Chornock}  \& {Filippenko}}{{Foley} et~al.}{2007}]{foley07}
{Foley} R.~J.,  {Smith} N.,  {Ganeshalingam} M.,  {Li} W.,  {Chornock} R.,
  {Filippenko} A.~V.,  2007, \mn@doi [\apjl] {10.1086/513145}, \href
  {http://adsabs.harvard.edu/abs/2007ApJ...657L.105F} {657, L105}

\bibitem[\protect\citeauthoryear{{Fox} et~al.,}{{Fox} et~al.}{2014}]{fox14}
{Fox} O.~D.,  et~al., 2014, \mn@doi [\apj] {10.1088/0004-637X/790/1/17}, \href
  {http://adsabs.harvard.edu/abs/2014ApJ...790...17F} {790, 17}

\bibitem[\protect\citeauthoryear{{Gies}, {Bagnuolo}, {Ferrara}, {Kaye},
  {Thaller}, {Penny}  \& {Peters}}{{Gies} et~al.}{1998}]{1998ApJ...493..440G}
{Gies} D.~R.,  {Bagnuolo} Jr. W.~G.,  {Ferrara} E.~C.,  {Kaye} A.~B.,
  {Thaller} M.~L.,  {Penny} L.~R.,   {Peters} G.~J.,  1998, \mn@doi [\apj]
  {10.1086/305113}, \href {http://adsabs.harvard.edu/abs/1998ApJ...493..440G}
  {493, 440}

\bibitem[\protect\citeauthoryear{{G{\"o}tberg}, {de Mink}  \&
  {Groh}}{{G{\"o}tberg} et~al.}{2017}]{2017A&A...608A..11G}
{G{\"o}tberg} Y.,  {de Mink} S.~E.,   {Groh} J.~H.,  2017, \mn@doi [\aap]
  {10.1051/0004-6361/201730472}, \href
  {http://adsabs.harvard.edu/abs/2017A%26A...608A..11G} {608, A11}

\bibitem[\protect\citeauthoryear{{Graczyk} et~al.,}{{Graczyk}
  et~al.}{2011}]{2011AcA....61..103G}
{Graczyk} D.,  et~al., 2011, \actaa, \href
  {http://adsabs.harvard.edu/abs/2011AcA....61..103G} {61, 103}

\bibitem[\protect\citeauthoryear{{Groh}, {Oliveira}  \& {Steiner}}{{Groh}
  et~al.}{2008}]{2008A&A...485..245G}
{Groh} J.~H.,  {Oliveira} A.~S.,   {Steiner} J.~E.,  2008, \mn@doi [\aap]
  {10.1051/0004-6361:200809511}, \href
  {http://adsabs.harvard.edu/abs/2008A%26A...485..245G} {485, 245}

\bibitem[\protect\citeauthoryear{{Groh}, {Meynet}, {Georgy}  \&
  {Ekstr{\"o}m}}{{Groh} et~al.}{2013}]{groh14}
{Groh} J.~H.,  {Meynet} G.,  {Georgy} C.,   {Ekstr{\"o}m} S.,  2013, \mn@doi
  [\aap] {10.1051/0004-6361/201321906}, \href
  {http://adsabs.harvard.edu/abs/2013A%26A...558A.131G} {558, A131}

\bibitem[\protect\citeauthoryear{{Hachinger}, {Mazzali}, {Taubenberger},
  {Hillebrandt}, {Nomoto}  \& {Sauer}}{{Hachinger} et~al.}{2012}]{haschinger12}
{Hachinger} S.,  {Mazzali} P.~A.,  {Taubenberger} S.,  {Hillebrandt} W.,
  {Nomoto} K.,   {Sauer} D.~N.,  2012, \mn@doi [\mnras]
  {10.1111/j.1365-2966.2012.20464.x}, \href
  {http://adsabs.harvard.edu/abs/2012MNRAS.422...70H} {422, 70}

\bibitem[\protect\citeauthoryear{{Hainich}, {Shenar}, {Sander}, {Hamann}  \&
  {Todt}}{{Hainich} et~al.}{2017}]{2017IAUS..329..171H}
{Hainich} R.,  {Shenar} T.,  {Sander} A.,  {Hamann} W.-R.,   {Todt} H.,  2017,
  in {Eldridge} J.~J.,  {Bray} J.~C.,  {McClelland} L.~A.~S.,   {Xiao} L.,
  eds,  IAU Symposium Vol. 329, The Lives and Death-Throes of Massive Stars. pp
  171--175 (\mn@eprint {arXiv} {1703.02060}),
  \mn@doi{10.1017/S1743921317002794}

\bibitem[\protect\citeauthoryear{{Hillier} \& {Miller}}{{Hillier} \&
  {Miller}}{1998}]{Hillier+1998}
{Hillier} D.~J.,  {Miller} D.~L.,  1998, \mn@doi [\apj] {10.1086/305350}, \href
  {http://adsabs.harvard.edu/abs/1998ApJ...496..407H} {496, 407}

\bibitem[\protect\citeauthoryear{{Hoogerwerf}, {de Bruijne}  \& {de
  Zeeuw}}{{Hoogerwerf} et~al.}{2001}]{Hoogerwerf+2001}
{Hoogerwerf} R.,  {de Bruijne} J.~H.~J.,   {de Zeeuw} P.~T.,  2001, \mn@doi
  [\aap] {10.1051/0004-6361:20000014}, \href
  {http://adsabs.harvard.edu/abs/2001A%26A...365...49H} {365, 49}

\bibitem[\protect\citeauthoryear{{Hosseinzadeh} et~al.,}{{Hosseinzadeh}
  et~al.}{2017}]{hoss17}
{Hosseinzadeh} G.,  et~al., 2017, \mn@doi [\apj] {10.3847/1538-4357/836/2/158},
  \href {http://adsabs.harvard.edu/abs/2017ApJ...836..158H} {836, 158}

\bibitem[\protect\citeauthoryear{{Hunter} et~al.,}{{Hunter}
  et~al.}{2009}]{Hunter+2009a}
{Hunter} I.,  et~al., 2009, \mn@doi [\aap] {10.1051/0004-6361/200809925}, \href
  {http://adsabs.harvard.edu/abs/2009A%26A...496..841H} {496, 841}

\bibitem[\protect\citeauthoryear{{Kim}, {Yoon}  \& {Koo}}{{Kim}
  et~al.}{2015}]{2015ApJ...809..131K}
{Kim} H.-J.,  {Yoon} S.-C.,   {Koo} B.-C.,  2015, \mn@doi [\apj]
  {10.1088/0004-637X/809/2/131}, \href
  {http://adsabs.harvard.edu/abs/2015ApJ...809..131K} {809, 131}

\bibitem[\protect\citeauthoryear{{Kobulnicky} et~al.,}{{Kobulnicky}
  et~al.}{2014}]{Kobulnicky+2014}
{Kobulnicky} H.~A.,  et~al., 2014, \mn@doi [\apjs]
  {10.1088/0067-0049/213/2/34}, \href
  {http://adsabs.harvard.edu/abs/2014ApJS..213...34K} {213, 34}

\bibitem[\protect\citeauthoryear{{Kurucz}}{{Kurucz}}{1992}]{Kurucz1992}
{Kurucz} R.~L.,  1992, in {Barbuy} B.,  {Renzini} A.,  eds,  IAU Symposium Vol.
  149, The Stellar Populations of Galaxies. p.~225

\bibitem[\protect\citeauthoryear{{Lyman}, {Bersier}, {James}, {Mazzali},
  {Eldridge}, {Fraser}  \& {Pian}}{{Lyman} et~al.}{2016}]{lyman16}
{Lyman} J.~D.,  {Bersier} D.,  {James} P.~A.,  {Mazzali} P.~A.,  {Eldridge}
  J.~J.,  {Fraser} M.,   {Pian} E.,  2016, \mn@doi [\mnras]
  {10.1093/mnras/stv2983}, \href
  {http://adsabs.harvard.edu/abs/2016MNRAS.457..328L} {457, 328}

\bibitem[\protect\citeauthoryear{{Maeder}}{{Maeder}}{1987}]{Maeder1987}
{Maeder} A.,  1987, \aap, \href
  {http://adsabs.harvard.edu/abs/1987A%26A...178..159M} {178, 159}

\bibitem[\protect\citeauthoryear{{Mandel} \& {de Mink}}{{Mandel} \& {de
  Mink}}{2016}]{Mandel+2016}
{Mandel} I.,  {de Mink} S.~E.,  2016, \mn@doi [\mnras] {10.1093/mnras/stw379},
  \href {http://adsabs.harvard.edu/abs/2016MNRAS.458.2634M} {458, 2634}

\bibitem[\protect\citeauthoryear{{Marchant}, {Langer}, {Podsiadlowski},
  {Tauris}  \& {Moriya}}{{Marchant} et~al.}{2016}]{2016A&A...588A..50M}
{Marchant} P.,  {Langer} N.,  {Podsiadlowski} P.,  {Tauris} T.~M.,   {Moriya}
  T.~J.,  2016, \mn@doi [\aap] {10.1051/0004-6361/201628133}, \href
  {http://adsabs.harvard.edu/abs/2016A%26A...588A..50M} {588, A50}

\bibitem[\protect\citeauthoryear{{Massey}}{{Massey}}{2003}]{massey03}
{Massey} P.,  2003, \mn@doi [\araa] {10.1146/annurev.astro.41.071601.170033},
  \href {http://adsabs.harvard.edu/abs/2003ARA%26A..41...15M} {41, 15}

\bibitem[\protect\citeauthoryear{{Massey}, {Neugent}, {Morrell}  \&
  {Hillier}}{{Massey} et~al.}{2014}]{massey14}
{Massey} P.,  {Neugent} K.~F.,  {Morrell} N.,   {Hillier} D.~J.,  2014, \mn@doi
  [\apj] {10.1088/0004-637X/788/1/83}, \href
  {http://adsabs.harvard.edu/abs/2014ApJ...788...83M} {788, 83}

\bibitem[\protect\citeauthoryear{{Massey}, {Neugent}  \& {Morrell}}{{Massey}
  et~al.}{2015}]{massey15}
{Massey} P.,  {Neugent} K.~F.,   {Morrell} N.,  2015, \mn@doi [\apj]
  {10.1088/0004-637X/807/1/81}, \href
  {http://adsabs.harvard.edu/abs/2015ApJ...807...81M} {807, 81}

\bibitem[\protect\citeauthoryear{{Massey}, {Neugent}  \& {Morrell}}{{Massey}
  et~al.}{2017}]{massey17}
{Massey} P.,  {Neugent} K.~F.,   {Morrell} N.,  2017, \mn@doi [\apj]
  {10.3847/1538-4357/aa5d17}, \href
  {http://adsabs.harvard.edu/abs/2017ApJ...837..122M} {837, 122}

\bibitem[\protect\citeauthoryear{{Maund}, {Smartt}, {Kudritzki},
  {Podsiadlowski}  \& {Gilmore}}{{Maund} et~al.}{2004}]{maund04}
{Maund} J.~R.,  {Smartt} S.~J.,  {Kudritzki} R.~P.,  {Podsiadlowski} P.,
  {Gilmore} G.~F.,  2004, \mn@doi [\nat] {10.1038/nature02161}, \href
  {http://adsabs.harvard.edu/abs/2004Natur.427..129M} {427, 129}

\bibitem[\protect\citeauthoryear{{Maund} et~al.,}{{Maund}
  et~al.}{2011}]{maund11}
{Maund} J.~R.,  et~al., 2011, \mn@doi [\apjl] {10.1088/2041-8205/739/2/L37},
  \href {http://adsabs.harvard.edu/abs/2011ApJ...739L..37M} {739, L37}

\bibitem[\protect\citeauthoryear{{Meynet} \& {Maeder}}{{Meynet} \&
  {Maeder}}{2003}]{mm03}
{Meynet} G.,  {Maeder} A.,  2003, \mn@doi [\aap] {10.1051/0004-6361:20030512},
  \href {http://adsabs.harvard.edu/abs/2003A%26A...404..975M} {404, 975}

\bibitem[\protect\citeauthoryear{{Meynet}, {Georgy}, {Hirschi}, {Maeder},
  {Massey}, {Przybilla}  \& {Nieva}}{{Meynet} et~al.}{2011}]{meynet11}
{Meynet} G.,  {Georgy} C.,  {Hirschi} R.,  {Maeder} A.,  {Massey} P.,
  {Przybilla} N.,   {Nieva} M.-F.,  2011, Bulletin de la Societe Royale des
  Sciences de Liege, \href {http://adsabs.harvard.edu/abs/2011BSRSL..80..266M}
  {80, 266}

\bibitem[\protect\citeauthoryear{{Moe} \& {Di Stefano}}{{Moe} \& {Di
  Stefano}}{2017}]{2016arXiv160605347M}
{Moe} M.,  {Di Stefano} R.,  2017, \mn@doi [\apjs] {10.3847/1538-4365/aa6fb6},
  \href {http://adsabs.harvard.edu/abs/2017ApJS..230...15M} {230, 15}

\bibitem[\protect\citeauthoryear{{Neugent}, {Massey}, {Skiff}  \&
  {Meynet}}{{Neugent} et~al.}{2012}]{2012ApJ...749..177N}
{Neugent} K.~F.,  {Massey} P.,  {Skiff} B.,   {Meynet} G.,  2012, \mn@doi
  [\apj] {10.1088/0004-637X/749/2/177}, \href
  {http://adsabs.harvard.edu/abs/2012ApJ...749..177N} {749, 177}

\bibitem[\protect\citeauthoryear{{Neugent}, {Massey}, {Hillier}  \&
  {Morrell}}{{Neugent} et~al.}{2016}]{neugent16}
{Neugent} K.~F.,  {Massey} P.,  {Hillier} D.~J.,   {Morrell} N.~I.,  2016, in
  {Hamann} W.-R.,  {Sander} A.,   {Todt} H.,  eds,  Vol. 517, (Potsdam:
  Verlag).

\bibitem[\protect\citeauthoryear{{Neugent}, {Massey}, {Hillier}  \&
  {Morrell}}{{Neugent} et~al.}{2017}]{neugent17}
{Neugent} K.~F.,  {Massey} P.,  {Hillier} D.~J.,   {Morrell} N.,  2017, \mn@doi
  [\apj] {10.3847/1538-4357/aa6e51}, \href
  {http://adsabs.harvard.edu/abs/2017ApJ...841...20N} {841, 20}

\bibitem[\protect\citeauthoryear{{Nieuwenhuijzen} \& {de
  Jager}}{{Nieuwenhuijzen} \& {de Jager}}{1990}]{Nieuwenhuijzen+1990}
{Nieuwenhuijzen} H.,  {de Jager} C.,  1990, \aap, \href
  {http://adsabs.harvard.edu/abs/1990A%26A...231..134N} {231, 134}

\bibitem[\protect\citeauthoryear{{Nugis} \& {Lamers}}{{Nugis} \&
  {Lamers}}{2000}]{Nugis+2000}
{Nugis} T.,  {Lamers} H.~J.~G.~L.~M.,  2000, \aap, \href
  {http://adsabs.harvard.edu/abs/2000A%26A...360..227N} {360, 227}

\bibitem[\protect\citeauthoryear{{{\"O}pik}}{{{\"O}pik}}{1924}]{opik24}
{{\"O}pik} E.,  1924, Publications of the Tartu Astrofizica Observatory, \href
  {http://adsabs.harvard.edu/abs/1924PTarO..25f...1O} {25}

\bibitem[\protect\citeauthoryear{{Packet}}{{Packet}}{1981}]{Packet1981}
{Packet} W.,  1981, \aap, \href
  {http://adsabs.harvard.edu/abs/1981A%26A...102...17P} {102, 17}

\bibitem[\protect\citeauthoryear{{Paczy{\'n}ski}}{{Paczy{\'n}ski}}{1971}]{paczynski71}
{Paczy{\'n}ski} B.,  1971, \mn@doi [\araa]
  {10.1146/annurev.aa.09.090171.001151}, \href
  {http://adsabs.harvard.edu/abs/1971ARA%26A...9..183P} {9, 183}

\bibitem[\protect\citeauthoryear{{Pastorello} et~al.,}{{Pastorello}
  et~al.}{2007}]{pastorello07}
{Pastorello} A.,  et~al., 2007, \mn@doi [\nat] {10.1038/nature05825}, \href
  {http://adsabs.harvard.edu/abs/2007Natur.447..829P} {447, 829}

\bibitem[\protect\citeauthoryear{{Pastorello} et~al.,}{{Pastorello}
  et~al.}{2008}]{pastorello08}
{Pastorello} A.,  et~al., 2008, \mn@doi [\mnras]
  {10.1111/j.1365-2966.2008.13602.x}, \href
  {http://adsabs.harvard.edu/abs/2008MNRAS.389..113P} {389, 113}

\bibitem[\protect\citeauthoryear{{Paxton}, {Bildsten}, {Dotter}, {Herwig},
  {Lesaffre}  \& {Timmes}}{{Paxton} et~al.}{2011}]{Paxton+2011}
{Paxton} B.,  {Bildsten} L.,  {Dotter} A.,  {Herwig} F.,  {Lesaffre} P.,
  {Timmes} F.,  2011, \mn@doi [\apjs] {10.1088/0067-0049/192/1/3}, \href
  {http://adsabs.harvard.edu/abs/2011ApJS..192....3P} {192, 3}

\bibitem[\protect\citeauthoryear{{Paxton} et~al.,}{{Paxton}
  et~al.}{2013}]{Paxton+2013}
{Paxton} B.,  et~al., 2013, \mn@doi [\apjs] {10.1088/0067-0049/208/1/4}, \href
  {http://adsabs.harvard.edu/abs/2013ApJS..208....4P} {208, 4}

\bibitem[\protect\citeauthoryear{{Paxton} et~al.,}{{Paxton}
  et~al.}{2015}]{2015ApJS..220...15P}
{Paxton} B.,  et~al., 2015, \mn@doi [\apjs] {10.1088/0067-0049/220/1/15}, \href
  {http://adsabs.harvard.edu/abs/2015ApJS..220...15P} {220, 15}

\bibitem[\protect\citeauthoryear{{Paxton} et~al.,}{{Paxton}
  et~al.}{2017}]{2017arXiv171008424P}
{Paxton} B.,  et~al., 2017, preprint, \href
  {http://adsabs.harvard.edu/abs/2017arXiv171008424P} {} (\mn@eprint {arXiv}
  {1710.08424})

\bibitem[\protect\citeauthoryear{{Peters}, {Gies}, {Grundstrom}  \&
  {McSwain}}{{Peters} et~al.}{2008}]{2008ApJ...686.1280P}
{Peters} G.~J.,  {Gies} D.~R.,  {Grundstrom} E.~D.,   {McSwain} M.~V.,  2008,
  \mn@doi [\apj] {10.1086/591145}, \href
  {http://adsabs.harvard.edu/abs/2008ApJ...686.1280P} {686, 1280}

\bibitem[\protect\citeauthoryear{{Peters}, {Pewett}, {Gies}, {Touhami}  \&
  {Grundstrom}}{{Peters} et~al.}{2013}]{peters13}
{Peters} G.~J.,  {Pewett} T.~D.,  {Gies} D.~R.,  {Touhami} Y.~N.,
  {Grundstrom} E.~D.,  2013, \mn@doi [\apj] {10.1088/0004-637X/765/1/2}, \href
  {http://adsabs.harvard.edu/abs/2013ApJ...765....2P} {765, 2}

\bibitem[\protect\citeauthoryear{{Peters}, {Wang}, {Gies}  \&
  {Grundstrom}}{{Peters} et~al.}{2016}]{2016ApJ...828...47P}
{Peters} G.~J.,  {Wang} L.,  {Gies} D.~R.,   {Grundstrom} E.~D.,  2016, \mn@doi
  [\apj] {10.3847/0004-637X/828/1/47}, \href
  {http://adsabs.harvard.edu/abs/2016ApJ...828...47P} {828, 47}

\bibitem[\protect\citeauthoryear{{Pols}}{{Pols}}{1994}]{Pols1994a}
{Pols} O.~R.,  1994, \aap, \href
  {http://adsabs.harvard.edu/abs/1994A%26A...290..119P} {290, 119}

\bibitem[\protect\citeauthoryear{{Renzo}, {Ott}, {Shore}  \& {de Mink}}{{Renzo}
  et~al.}{2017}]{renzo17}
{Renzo} M.,  {Ott} C.~D.,  {Shore} S.~N.,   {de Mink} S.~E.,  2017, \mn@doi
  [\aap] {10.1051/0004-6361/201730698}, \href
  {http://adsabs.harvard.edu/abs/2017A%26A...603A.118R} {603, A118}

\bibitem[\protect\citeauthoryear{{Sana} et~al.,}{{Sana} et~al.}{2012}]{sana12}
{Sana} H.,  et~al., 2012, \mn@doi [Science] {10.1126/science.1223344}, \href
  {http://adsabs.harvard.edu/abs/2012Sci...337..444S} {337, 444}

\bibitem[\protect\citeauthoryear{{Sanders} et~al.,}{{Sanders}
  et~al.}{2013}]{sanders13}
{Sanders} N.~E.,  et~al., 2013, \mn@doi [\apj] {10.1088/0004-637X/769/1/39},
  \href {http://adsabs.harvard.edu/abs/2013ApJ...769...39S} {769, 39}

\bibitem[\protect\citeauthoryear{{Schneider}, {Izzard}, {Langer}  \& {de
  Mink}}{{Schneider} et~al.}{2015}]{Schneider+2015}
{Schneider} F.~R.~N.,  {Izzard} R.~G.,  {Langer} N.,   {de Mink} S.~E.,  2015,
  \mn@doi [\apj] {10.1088/0004-637X/805/1/20}, \href
  {http://adsabs.harvard.edu/abs/2015ApJ...805...20S} {805, 20}

\bibitem[\protect\citeauthoryear{{Shivvers} et~al.,}{{Shivvers}
  et~al.}{2017}]{shivvers17}
{Shivvers} I.,  et~al., 2017, \mn@doi [\pasp] {10.1088/1538-3873/aa54a6}, \href
  {http://adsabs.harvard.edu/abs/2017PASP..129e4201S} {129, 054201}

\bibitem[\protect\citeauthoryear{{Smartt}}{{Smartt}}{2009}]{smartt09}
{Smartt} S.~J.,  2009, \mn@doi [\araa] {10.1146/annurev-astro-082708-101737},
  \href {http://adsabs.harvard.edu/abs/2009ARA%26A..47...63S} {47, 63}

\bibitem[\protect\citeauthoryear{{Smith}}{{Smith}}{2014}]{smith14}
{Smith} N.,  2014, \mn@doi [\araa] {10.1146/annurev-astro-081913-040025}, \href
  {http://adsabs.harvard.edu/abs/2014ARA%26A..52..487S} {52, 487}

\bibitem[\protect\citeauthoryear{{Smith}}{{Smith}}{2016}]{smith16}
{Smith} N.,  2016, \mn@doi [\mnras] {10.1093/mnras/stw1533}, \href
  {http://adsabs.harvard.edu/abs/2016MNRAS.461.3353S} {461, 3353}

\bibitem[\protect\citeauthoryear{{Smith} \& {Conti}}{{Smith} \&
  {Conti}}{2008}]{sc08}
{Smith} N.,  {Conti} P.~S.,  2008, \mn@doi [\apj] {10.1086/586885}, \href
  {http://adsabs.harvard.edu/abs/2008ApJ...679.1467S} {679, 1467}

\bibitem[\protect\citeauthoryear{{Smith} \& {Owocki}}{{Smith} \&
  {Owocki}}{2006}]{so06}
{Smith} N.,  {Owocki} S.~P.,  2006, \mn@doi [\apjl] {10.1086/506523}, \href
  {http://adsabs.harvard.edu/abs/2006ApJ...645L..45S} {645, L45}

\bibitem[\protect\citeauthoryear{{Smith} \& {Tombleson}}{{Smith} \&
  {Tombleson}}{2015}]{st15}
{Smith} N.,  {Tombleson} R.,  2015, \mn@doi [\mnras] {10.1093/mnras/stu2430},
  \href {http://adsabs.harvard.edu/abs/2015MNRAS.447..598S} {447, 598}

\bibitem[\protect\citeauthoryear{{Smith}, {Li}, {Filippenko}  \&
  {Chornock}}{{Smith} et~al.}{2011}]{smith11}
{Smith} N.,  {Li} W.,  {Filippenko} A.~V.,   {Chornock} R.,  2011, \mn@doi
  [\mnras] {10.1111/j.1365-2966.2011.17229.x}, \href
  {http://adsabs.harvard.edu/abs/2011MNRAS.412.1522S} {412, 1522}

\bibitem[\protect\citeauthoryear{{Smith}, {Andrews}  \& {Mauerhan}}{{Smith}
  et~al.}{2016}]{smith+16}
{Smith} N.,  {Andrews} J.~E.,   {Mauerhan} J.~C.,  2016, \mn@doi [\mnras]
  {10.1093/mnras/stw2190}, \href
  {http://adsabs.harvard.edu/abs/2016MNRAS.463.2904S} {463, 2904}

\bibitem[\protect\citeauthoryear{{Steiner} \& {Oliveira}}{{Steiner} \&
  {Oliveira}}{2005}]{2005A&A...444..895S}
{Steiner} J.~E.,  {Oliveira} A.~S.,  2005, \mn@doi [\aap]
  {10.1051/0004-6361:20052782}, \href
  {http://adsabs.harvard.edu/abs/2005A%26A...444..895S} {444, 895}

\bibitem[\protect\citeauthoryear{{Van Dyk} et~al.,}{{Van Dyk}
  et~al.}{2011}]{svd11}
{Van Dyk} S.~D.,  et~al., 2011, \mn@doi [\apjl] {10.1088/2041-8205/741/2/L28},
  \href {http://adsabs.harvard.edu/abs/2011ApJ...741L..28V} {741, L28}

\bibitem[\protect\citeauthoryear{{Van Dyk} et~al.,}{{Van Dyk}
  et~al.}{2014}]{svd14}
{Van Dyk} S.~D.,  et~al., 2014, \mn@doi [\aj] {10.1088/0004-6256/147/2/37},
  \href {http://adsabs.harvard.edu/abs/2014AJ....147...37V} {147, 37}

\bibitem[\protect\citeauthoryear{{Van Dyk}, {de Mink}  \& {Zapartas}}{{Van Dyk}
  et~al.}{2016}]{svd16}
{Van Dyk} S.~D.,  {de Mink} S.~E.,   {Zapartas} E.,  2016, \mn@doi [\apj]
  {10.3847/0004-637X/818/1/75}, \href
  {http://adsabs.harvard.edu/abs/2016ApJ...818...75V} {818, 75}

\bibitem[\protect\citeauthoryear{{Vink}}{{Vink}}{2017}]{2017A&A...607L...8V}
{Vink} J.~S.,  2017, \mn@doi [\aap] {10.1051/0004-6361/201731902}, \href
  {http://adsabs.harvard.edu/abs/2017A%26A...607L...8V} {607, L8}

\bibitem[\protect\citeauthoryear{{Vink}, {de Koter}  \& {Lamers}}{{Vink}
  et~al.}{2000}]{Vink+2000}
{Vink} J.~S.,  {de Koter} A.,   {Lamers} H.~J.~G.~L.~M.,  2000, \aap, \href
  {http://adsabs.harvard.edu/abs/2000A%26A...362..295V} {362, 295}

\bibitem[\protect\citeauthoryear{{Vink}, {de Koter}  \& {Lamers}}{{Vink}
  et~al.}{2001}]{Vink+2001}
{Vink} J.~S.,  {de Koter} A.,   {Lamers} H.~J.~G.~L.~M.,  2001, \mn@doi [\aap]
  {10.1051/0004-6361:20010127}, \href
  {http://adsabs.harvard.edu/abs/2001A%26A...369..574V} {369, 574}

\bibitem[\protect\citeauthoryear{{Wang}, {Gies}  \& {Peters}}{{Wang}
  et~al.}{2017}]{2017ApJ...843...60W}
{Wang} L.,  {Gies} D.~R.,   {Peters} G.~J.,  2017, \mn@doi [\apj]
  {10.3847/1538-4357/aa740a}, \href
  {http://adsabs.harvard.edu/abs/2017ApJ...843...60W} {843, 60}

\bibitem[\protect\citeauthoryear{{Yoon} \& {Langer}}{{Yoon} \&
  {Langer}}{2005}]{Yoon+2005}
{Yoon} S.-C.,  {Langer} N.,  2005, \mn@doi [\aap] {10.1051/0004-6361:20054030},
  \href {http://adsabs.harvard.edu/abs/2005A%26A...443..643Y} {443, 643}

\bibitem[\protect\citeauthoryear{{Yoon}, {Langer}  \& {Norman}}{{Yoon}
  et~al.}{2006}]{Yoon+2006}
{Yoon} S.-C.,  {Langer} N.,   {Norman} C.,  2006, \mn@doi [\aap]
  {10.1051/0004-6361:20065912}, \href
  {http://adsabs.harvard.edu/abs/2006A%26A...460..199Y} {460, 199}

\bibitem[\protect\citeauthoryear{{Yoon}, {Dessart}  \& {Clocchiatti}}{{Yoon}
  et~al.}{2017}]{yoon17}
{Yoon} S.-C.,  {Dessart} L.,   {Clocchiatti} A.,  2017, \mn@doi [\apj]
  {10.3847/1538-4357/aa6afe}, \href
  {http://adsabs.harvard.edu/abs/2017ApJ...840...10Y} {840, 10}

\bibitem[\protect\citeauthoryear{{Zapartas} \& {et al.}}{{Zapartas} \& {et
  al.}}{2017}]{zapartas17}
{Zapartas} E.,  {et al.} 2017, \apj

\bibitem[\protect\citeauthoryear{{de Jager}, {Nieuwenhuijzen}  \& {van der
  Hucht}}{{de Jager} et~al.}{1988}]{de-Jager+1988}
{de Jager} C.,  {Nieuwenhuijzen} H.,   {van der Hucht} K.~A.,  1988, \aaps,
  \href {http://adsabs.harvard.edu/abs/1988A%26AS...72..259D} {72, 259}

\bibitem[\protect\citeauthoryear{{de Mink}, {Langer}, {Izzard}, {Sana}  \& {de
  Koter}}{{de Mink} et~al.}{2013}]{de-Mink+2013}
{de Mink} S.~E.,  {Langer} N.,  {Izzard} R.~G.,  {Sana} H.,   {de Koter} A.,
  2013, \mn@doi [\apj] {10.1088/0004-637X/764/2/166}, \href
  {http://adsabs.harvard.edu/abs/2013ApJ...764..166D} {764, 166}

\makeatother
\end{thebibliography}

\end{document}